\documentclass[natbib2009,numberedappendix]{emulateapj}
\usepackage{natbib}
\usepackage{amsmath}
\usepackage{multirow}
\usepackage{color}

\newcommand{\cii}{\mbox{[C{\small II}]}}
\newcommand{\oi}{\mbox{[O{\small I}]}}

\shorttitle{\cii\ 158~$\mu$m Emission as a Star Formation Tracer}
\shortauthors{Herrera-Camus et al.}

\begin{document}

\title{\cii\ 158~$\mu$m Emission as a Star Formation Tracer}


\author{R.~Herrera-Camus\altaffilmark{1,2}}
\author{A.~D.~Bolatto\altaffilmark{1,2}}
\author{M.~G.~Wolfire\altaffilmark{1}}
\author{J.~D.~Smith\altaffilmark{3}} 
\author{K.~V.~Croxall\altaffilmark{4}}
\author{R.~C.~Kennicutt\altaffilmark{5}}
\author{D.~Calzetti\altaffilmark{6}}
\author{G.~Helou\altaffilmark{7}}
\author{F.~Walter\altaffilmark{8}}
\author{A.~K.~Leroy\altaffilmark{9}}
\author{B.~Draine\altaffilmark{10}}
\author{B.~R.~Brandl\altaffilmark{11}}
\author{L.~Armus\altaffilmark{12}}
\author{K.~M.~Sandstrom\altaffilmark{13}}
\author{D.~A.~Dale\altaffilmark{14}}
\author{G.~Aniano\altaffilmark{15}}
\author{S.~E.~Meidt\altaffilmark{8}}
\author{M.~Boquien\altaffilmark{5}}
\author{L.~K.~Hunt\altaffilmark{16}}
\author{M.~Galametz\altaffilmark{17}}
\author{F.~S.~Tabatabaei\altaffilmark{8}}
\author{E.~J.~Murphy\altaffilmark{7}}
\author{P.~Appleton\altaffilmark{18}}
\author{H.~Roussel\altaffilmark{19}}
\author{C.~Engelbracht\altaffilmark{20,21,$^{\star}$}}
\author{P.~Beirao\altaffilmark{22}}

\altaffiltext{1}{Department of Astronomy, University of Maryland, College Park, MD 20742, USA.}
\altaffiltext{2}{Laboratory for Millimeter-wave Astronomy, University of Maryland, College Park, MD 20742, USA}
\altaffiltext{3}{Department of Physics and Astronomy, University of Toledo, 2801 West Bancroft Street, Toledo, OH 43606, USA}
\altaffiltext{4}{Department of Astronomy, The Ohio State University, 4051 McPherson Laboratory, 140 West 18th Avenue, Columbus, OH 43210, USA}
\altaffiltext{5}{Institute of Astronomy, University of Cambridge, Madingley Road, Cambridge CB3 0HA, UK}
\altaffiltext{6}{Department of Astronomy, University of Massachusetts, Amherst, MA 01003, USA}
\altaffiltext{7}{Infrared Processing and Analysis Center, California Institute of Technology, Pasadena, CA 91125, USA}
\altaffiltext{8}{Max-Planck-Institut f{\"u}r Astronomie, K{\"o}nigstuhl 17, D-69117 Heidelberg, Germany}
\altaffiltext{9}{National Radio Astronomy Observatory, 520 Edgemont Road, Charlottesville, VA 22903, USA}
\altaffiltext{10}{Department of Astrophysical Sciences, Princeton University, Princeton, NJ 08544, USA}
\altaffiltext{11}{Leiden Observatory, Leiden University, P.O. Box 9513, 2300-RA Leiden, The Netherlands}
\altaffiltext{12}{Spitzer Science Center, California Institute of Technology, MC 314-6, Pasadena, CA 91125, USA}
\altaffiltext{13}{Steward Observatory, University of Arizona, 933 North Cherry Avenue, Tucson, AZ 85721, USA}
\altaffiltext{14}{Department of Physics and Astronomy, University of Wyoming, Laramie, WY 82071, USA}
\altaffiltext{15}{Institut d'Astrophysique Spatiale, CNRS (UMR8617) Universit\'e
Paris-Sud 11, Batiment 121, Orsay, France}
\altaffiltext{16}{INAF-Osservatorio Astrofisico di Arcetri, Largo E. Fermi 5, I-50125 Firenze, Italy}
\altaffiltext{17}{European Southern Observatory, Karl Schwarzschild Str. 2, 85748 Garching, Germany}
\altaffiltext{18}{NASA Herschel Science Center, IPAC, California Institute of Technology, Pasadena, CA 91125, USA}
\altaffiltext{19}{Institut d'Astrophysique de Paris, Universit\'e Pierre et Marie Curie, CNRS UMR 7095, 75014 Paris, France}
\altaffiltext{20}{Steward Observatory, University of Arizona, Tucson, AZ 85721, USA}
\altaffiltext{21}{Raytheon Company, 1151 East Hermans Road, Tucson, AZ 85756, USA}
\altaffiltext{$^{\star}$}{Deceased January 8, 2014}
\altaffiltext{22}{Observatoire de Paris, 61 avenue de l'Observatoire, Paris F-75014, France}

\begin{abstract}
The \cii~157.74~$\mu$m transition is the dominant coolant of the neutral interstellar gas, and has great potential as a star formation rate (SFR) tracer. Using the {\em Herschel} KINGFISH sample of 46 nearby galaxies, we investigate the relation of \cii\ surface brightness and luminosity with SFR. We conclude that \cii\ can be used for measurements of SFR on both global and kiloparsec scales in normal star-forming galaxies in the absence of strong active galactic nuclei (AGN). The uncertainty of the $\Sigma_{\rm [CII]}-\Sigma_{\rm SFR}$ calibration is $\pm$0.21~dex. The main source of scatter in the correlation is associated with regions that exhibit warm IR colors, and we provide an adjustment based on IR color that reduces the scatter. We show that the color-adjusted $\Sigma_{\rm[CII]}-\Sigma_{\rm SFR}$ correlation is valid over almost 5 orders of magnitude in $\Sigma_{\rm SFR}$, holding for both normal star-forming galaxies and non-AGN luminous infrared galaxies. Using \cii\ luminosity instead of surface brightness to estimate SFR suffers from worse systematics, frequently underpredicting SFR in luminous infrared galaxies even after IR color adjustment (although this depends on the SFR measure employed). We suspect that surface brightness relations are better behaved than the luminosity relations because the former are more closely related to the local far-UV field strength, most likely the main parameter controlling the efficiency of the conversion of far-UV radiation into gas heating.  A simple model based on {\it Starburst99} population-synthesis code to connect SFR to \cii\ finds that heating efficiencies are $1\%-3\%$ in normal galaxies.
\end{abstract}


\section{{\bf Introduction}} 

The \cii\ 157.74~$\mu$m fine-structure transition ($^{2}P_{3/2}-^{2}P_{1/2}$) is one of the brightest emission lines in star-forming galaxies \citep{rhc_stacey91b,rhc_stacey10} and a major coolant for the neutral atomic gas \citep{rhc_wolfire03}. In this phase of the interstellar medium (ISM), far-ultraviolet (FUV) photons produced by O and B stars heat the gas via the photoelectric effect on small dust grains and polycyclic aromatic hydrocarbons \citep[PAHs;][]{rhc_helou01}. The ejected photoelectrons are thermalized and heat the gas. Neutral collisions marginally dominate the excitation of the fine-structure level of singly ionized carbon atoms and the gas cools by emission of \cii\ 158~$\mu$m photons. This chain of events provides a link between the star formation activity and the \cii\ emission: if the gas is in thermal balance, and it is cooled mainly by 158~$\mu$m emission, the \cii\ line measures the total energy that is put into the gas by star formation activity.

The ionization potential of neutral carbon is 11.3~eV, so ionized carbon (C$^{+}$) can be found in phases of the ISM where hydrogen is in molecular, neutral atomic or ionized form. C$^{+}$ can be excited by collisions with electrons (e$^{-}$), hydrogen atoms (H) and molecules (H$_{2}$). Assuming collisional excitation, the \cii\ integrated line intensity ($I_{\rm\cii}$) in the optically thin limit is \citep{rhc_crawford85}

\begin{equation}
I_{\rm[CII]} = 2.3 \times10^{-24} \bigg[\frac{2e^{-91.2/T}}{1+2e^{-91.2/T}+n_{\rm crit}/n}\bigg] N_{\rm C^{+}},
\end{equation}

\noindent where $I_{\rm[CII]}$ is in units of W~m$^{-2}$~sr$^{-1}$, $T$ is the kinetic temperature in K, $n$ is the volume density of the collisional partner (H, H$_{2}$ or e$^{-}$) in cm$^{-3}$,  $N_{\rm C^{+}}$ is the column density of C$^{+}$ in cm$^{-2}$ and $n_{\rm crit}$ is the critical density for collisions with a given partner in cm$^{-3}$. The latter is a function of temperature. For a typical cold neutral medium (CNM) temperature of $T\approx100$~K \citep{rhc_heiles03,rhc_wolfire03}, the critical density for collisions with e$^{-}$ and H is $9$~cm$^{-3}$ and $3,000$~cm$^{-3}$ respectively \citep{rhc_goldsmith12}. Typical volume densities of H atoms in the CNM are n$_{\rm H}\approx50$~cm$^{-3}$ ($\ll n_{\rm crit}({\rm H})$) and the fractional ionization n$_{\rm e^{-}}/{\rm n}_{\rm H}\lesssim10^{-3}$, so in this phase collisions with H atoms dominate the C$^{+}$ excitation. In the dense gas interface between molecular clouds and HII regions -- also known as phototodissociation regions (PDRs) -- the excitation is dominated by collisions with molecular hydrogen. At a gas temperature of  100~K, the critical density for collisions with H$_{2}$ is 6,100~cm$^{-3}$ \citep{rhc_goldsmith12}. In the warm ionized medium (WIM), for a characteristic temperature of $T\approx8,000$~K \citep{rhc_mckee77,rhc_haffner99}, the critical density for collisions with e$^{-}$ is 44~cm$^{-3}$ \citep{rhc_goldsmith12}. For a range of electron densities in the WIM of $\sim0.08-0.4$~cm$^{-3}$ \citep{rhc_haffner09,rhc_velusamy12}, collisions with e$^{-}$ are responsible for the excitation of C$^{+}$.

The multiphase contribution to the \cii\ emission includes the CNM, PDRs, HII regions and the WIM \citep{rhc_stacey85,rhc_shibai91,rhc_bennett94,rhc_stacey10,rhc_pineda13}. The individual contribution of each one of these ISM components to the total \cii\ luminosity is still a matter of study and depends on the nature of the object, location and resolution of the observations. In the Galactic plane, early observations by the COsmic Background Explorer (COBE) show that the \cii\ emission tends to follow the spiral arms and peaks at the molecular ring \citep{rhc_wright91,rhc_bennett94}. More recently, \cite{rhc_pineda13}, based on the {\it Herschel}/HIFI project ``Galactic Observations of Terahertz C+" (GOT C+), also find that in the plane of the galaxy the \cii\ emission is mostly associated with the spiral arms, with dense PDRs as the main source of the total \cii\ emission ($\sim47$\%), followed by atomic gas ($\sim21$\%) and small contributions from the ionized gas ($\sim4$\%). In low metallicity galaxies, the PDR contribution to the \cii\ emission can be dominant \citep[80\% in IC~10,][]{rhc_madden97} or small \citep[10\% in Haro~11,][]{rhc_cormier12}. Moving to higher redshifts, \cite{rhc_stacey10} find that for 3 starburst systems in the redshift range $z\sim1-2$ the origin of the \cii\ emission is also dominated by the PDR component, with the ratio of the \cii\ to the FIR emission similar to what is measured in nearby starburst galaxies.

Previous studies have searched for the connection between \cii\ emission and star formation activity. One of the first surveys of nearby, gas rich spirals observed in the \cii\ transition was done by \cite{rhc_stacey91b} using the Kuiper Airbone Observatory (KAO). They find that the integrated \cii/$^{12}$CO$(1-0)$ line intensity ratio for starburst nuclei is similar to the ratio measured in Galactic OB star-forming regions. They also measure ratios a factor $\sim$3 smaller in non-starburst systems and therefore, proposed to use this ratio as a tool to characterize the star formation activity. With the advent of the {\it Infrared Space Observatory} (ISO), \cite{rhc_boselli02} derive one of the first calibrations of the star formation rate (SFR) based on the \cii\ luminosity ($L_{\rm[CII]}$). For a sample of 22 late-type galaxies including galaxies from the Virgo cluster and M82, they find a nonlinear relationship between H$\alpha$ and \cii\ global luminosities ($L_{\text{H}\alpha} \propto L_{\rm[CII]}^{0.79}$) with a dispersion of at least a factor of $\sim3$. Also using ISO \cii\ observations, \cite{rhc_delooze11} find a nearly linear correlation between SFR(FUV+24~$\mu$m) and $L_{\rm[CII]}$ with a dispersion of $\sim0.3$~dex for a sample of 24 local, star-forming galaxies. 

More evidence in favor of \cii\ emission as a star formation tracer comes from {\it Herschel} observations. \cite{rhc_mookerjea11} find that the \cii\ emission in the M33 H{\small II} region, BCLMP~302, strongly correlates with H$\alpha$ and dust continuum emission on scales of $\sim$50~pc. More recently, \cite{rhc_sargsyan12}, \cite{rhc_delooze14} and \cite{rhc_pineda14} explore the \cii--SFR connection in luminous infrared galaxies (LIRGs, $L_{\rm IR}>10^{11}~L_\odot$), dwarf galaxies and the Milky Way, respectively. \cite{rhc_sargsyan12} find a linear relationship between the SFR(FIR) and $L_{\rm[CII]}$ for a sample of 24 LIRGs. \cite{rhc_delooze14}, using the Dwarf Galaxy Survey \citep{rhc_madden13}, conclude that the \oi~63~$\mu$m line is a better SFR tracer than \cii\ in low metallicity galaxies. \cite{rhc_pineda14} find that \cii\ emission emerging from different phases of the ISM in the Milky Way correlates well with SFR at Galactic scales. 

The \cii\ transition presents many advantages as a SFR indicator. Among these: (1) it is a very bright line, with luminosities tipically $\sim0.1-1\%$ of the FIR luminosity; (2) it is practically unaffected by extinction -- possible exceptions include edge-on galaxies \citep{rhc_heiles94} and extreme starbursts \citep{rhc_luhman98}; (3) it can be used to study star-forming galaxies at redshifts $z\gtrsim1$ using ground based-observatories like the Atacama Large Millimeter Array (ALMA) (e.g. see Figure~\ref{cii_alma}). For many of these high redshift objects, the \cii\ luminosity might be one of the few available tools to measure SFRs. 

The so called ``\cii\ deficit" is the most important potential limitation for using \cii\ as a SFR indicator. Observed in luminous and ultraluminous infrared galaxies \citep{rhc_malhotra97,rhc_malhotra01,rhc_brauher08,rhc_gracia-carpio11,rhc_diaz-santos13} and nearby galaxies \citep{rhc_beirao12, rhc_croxall12}, the ``\cii\ deficit" corresponds to lower \cii\ to FIR ratios measured as a function of increasingly warm infrared color. Several explanations for the observed low ratio of \cii\ to FIR have been proposed over the years. Some of these explanations may account only for a small subset of the low ratios: \cii\ self absorption, high dust extinction, softer UV radiation coming from older stellar populations \citep[see][for more discussion of these scenarios]{rhc_malhotra01}. Other alternatives seem to be applicable to a larger range of environments: (1) charging of the dust grains: at high radiation fields, the dust grains become positively charged \citep{rhc_malhotra97,rhc_croxall12}. A higher charge implies a higher Coulomb barrier for the photoelectrons to overcome; as a result the photoelectric heating efficiency drops. (2) \oi\ as an additional cooling channel: if the FUV radiation field and the density of the atomic gas increases above the critical density for collisional excitation with H atoms ($n_\text{crit}\sim10^{3}$~cm$^{-3}$), then collisional de-excitations start to suppress the \cii\ emission and the contribution to the cooling by the \oi~63~$\mu$m line ($n_{\rm crit}\sim10^{5}$~cm$^{-3}$) becomes dominant. (3) High ionization parameter \citep{rhc_gracia-carpio11}: in HII regions with high ionization parameter, a larger fraction of the non-ionizing stellar UV is absorbed by dust in the HII region, and thus a smaller fraction of the UV photons are available to heat the neutral gas.

The goal of this paper is to derive an accurate \cii--SFR calibration for normal galaxies, obtain a deeper understanding of the origin of the \cii--SFR correlation, and identify the limits of applicability of the calibration. To do this, we use a large sample of resolved extragalactic regions -- with a median size of $\sim$0.5~kpc -- selected from 46 nearby galaxies that are part of the KINGFISH sample of galaxies \citep[][Key Insights on Nearby Galaxies: A Far-Infrared Survey with Herschel]{rhc_kennicutt11}. This, combined with the wealth of ancillary data available -- such as IR, H$\alpha$ and FUV observations -- allow us to probe different timescales and environments associated with the star formation activity. 

This paper is organized as follows. In \S2 we describe the KINGFISH sample and the supplementary data. In \S3 we present the correlations between \cii\ and SFR estimated from 24~$\mu$m, total infrared (TIR), H$\alpha$ and FUV data. We also describe how we removed the cirrus and AGN contributions to the 24~$\mu$m emission. In \S4  we analyze the scatter of the \cii--SFR correlation in terms of the IR color and other properties of the ISM derived from the \cite{rhc_draine07} model. We also compare our calibration to previous \cii--SFR calibrations derived based on ISO and {\it Herschel} samples. Finally, we use the {\it Starburst99} code to analyze the scatter in terms of a combination of the duration of the star formation activity and the photoelectric heating efficiency of the dust grains.

\begin{figure*}
\epsscale{1}
\plotone{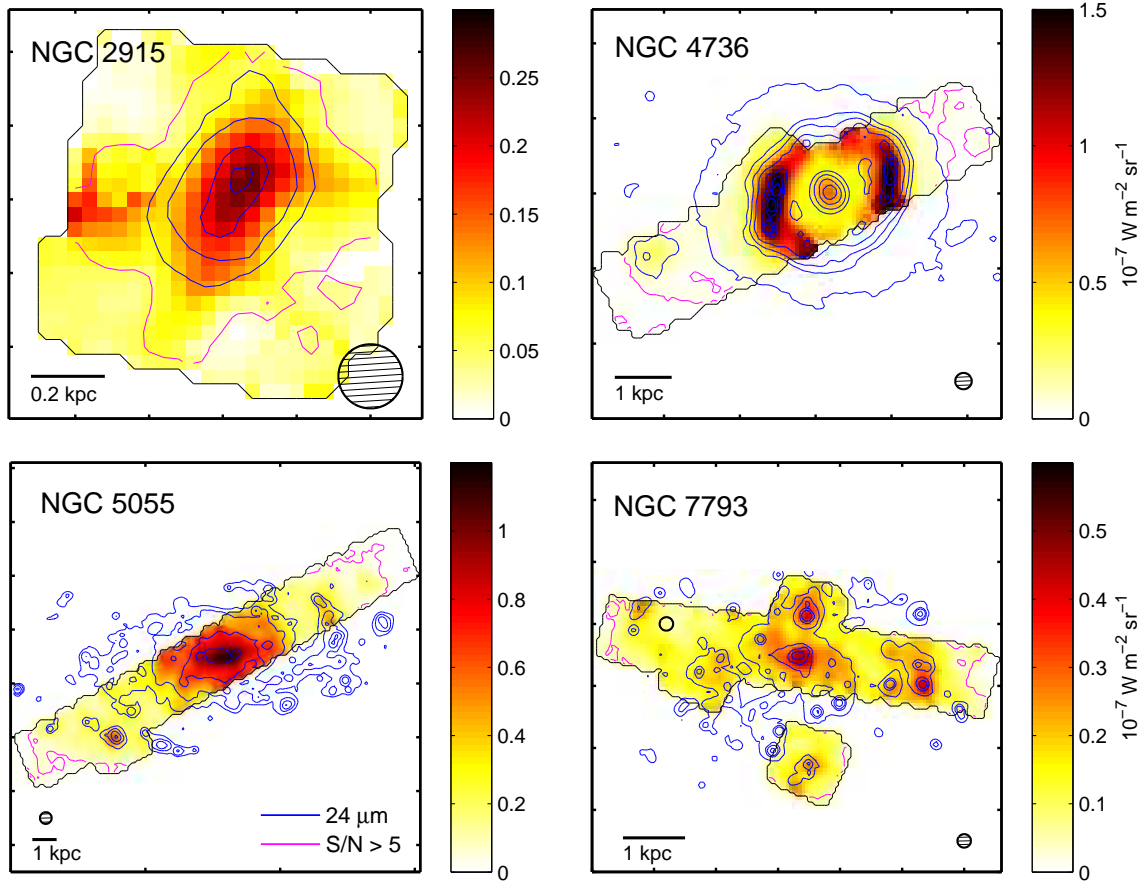}
\caption{PACS [CII] 158~$\mu$m images for four galaxies selected from the KINGFISH sample. The black contours delineate the areas which have [CII] data. The physical scale and the size of the $\sim12\arcsec$ beam are shown in the corners. The color scale shows the [CII] surface brightness in units of W~m$^{-2}$~sr$^{-1}$ and the blue contours show the 24~$\mu$m dust continuum emission.  At this spatial scale, there is a very good agreement between the [CII] line and the 24~$\mu$m continuum emission. The area enclosed by the magenta contours corresponds to the parts of the galaxy where the S/N of the [CII] emission is greater than 3. The four galaxies in the image are: ({\it top-left}) NGC 2915, a blue compact dwarf galaxy and one of the four systems in our sample with metallicity 12+log(O/H)~$<8.1$. This galaxy also has the lowest TIR luminosity of the sample ($L_{\rm TIR}=3.9\times10^{7}$~L$_{\odot}$). ({\it top-right}) NGC 4736, an early-type spiral galaxy with a circumnuclear ring traced by both, \cii\ and 24~$\mu$m emission. ({\it bottom-left}) NGC~5055, a spiral (SAbc) galaxy. The detected [CII] emission along the disk covers more than one order of magnitude in surface brightness. ({\it bottom-right}) NGC~7703, a flocculent spiral with [CII] emission mapped using a strip section and one extranuclear regions.\label{panel_galaxies}}
\end{figure*}

\section{{\bf Main Sample Description}} 

We focus our study on 46 galaxies from the KINGFISH sample \citep{rhc_kennicutt11}. KINGFISH combines deep {\it Herschel} infrared imaging with spectroscopy of the key interstellar medium diagnostic lines: \cii\ 158~$\mu$m, [NII]~122~$\mu$m \& 205~$\mu$m, \oi~63~$\mu$m and [OIII]~88~$\mu$m. Our spectroscopic sample includes 40 spiral galaxies that encompass the full range of late-type morphologies, as well as 4 irregulars (Holmberg~II, NGC~2915, NGC~3077 and NGC~5408), and 2 ellipticals (NGC ~855 and NGC~3265). 

There are eight other KINGFISH galaxies with spectroscopic data available that we do not include in this study.  These are: NGC~1266, NGC~1316, NGC~1097, NGC~1377, NGC~1404, NGC~4594, NGC~4631 and NGC~4559. The reasons why we exclude these galaxies are presented in Appendix~A.

Our sample spans more than three orders of magnitude in total infrared luminosity ($L_\text{TIR} \sim 10^{7.6} - 10^{11} L_{\odot}$) and about one order of magnitude in distance ($D \sim 2.8 - 26.5$~Mpc). The sample also covers a metallicity range of $12 + \text{log(O/H)} \sim 7.72 - 8.77$, measured by \cite{rhc_moustakas10} using the \cite{rhc_kobulnicky04} calibration. The beam size of the \cii\ 158~$\mu$m data is $\sim12''$; given the range of distances, our sample covers a range of spatial resolutions that goes from $\sim$0.2~kpc for IC~2574 to $\sim$1.5~kpc for NGC~5713, with a median value of $0.6\pm0.3$~kpc. In order to allow comparison between regions with different sizes, we report our measurements as luminosities per unit of physical area (surface brightness or luminosity surface density).

The FIR spectroscopic line observations were carried out with the Photodetector Array Camera \& Spectrometer (PACS) on board {\it Herschel} as part of the {\it Herschel} key program KINGFISH . The spectral observations were reduced using the {\it Herschel} Interactive Processing Environment (HIPE) version~11.0. The reduced cubes were then processed to obtain zero, first and second moment maps. For a detailed description on the reduction and processing of the KINGFISH FIR spectral maps we refer to \cite{rhc_croxall13}. About half of the images consists of strip maps, in some cases including extranuclear regions; the other half corresponds to rectangular regions centered on the nucleus of the galaxy.

The work in this paper is based on the \cii\ line, the brightest emission line in our sample. About 70\% of the galaxies in our sample show \cii\ line emission above the 3$\sigma$ level in at least 70\% of the map. Figure \ref{panel_galaxies} shows \cii\ surface brightness maps and 24~$\mu$m continuum contours for four KINGFISH galaxies: NGC 2915, irregular low metallicity galaxy with the lowest total infrared (TIR) luminosity in our sample; NGC 4736, spiral galaxy with a well defined circumnuclear ring visible in both, \cii\ and 24~$\mu$m emission; NGC 5055 and NGC 7793, two flocculent spiral systems with extended \cii\ emission detected across the disk and extranuclear regions. At this spatial scale, there is a very good agreement between the \cii\ line and the 24~$\mu$m dust continuum emission, a well characterized indicator of SFR.

In order to study the reliability of the \cii\ 158~$\mu$m line emission as a SFR tracer, it is crucial to combine this FIR line with archival data that provide information about the dust, gas and young stellar population. In Appendix~B we describe the supplementary data used in our analysis, which includes ground-based H$\alpha$, GALEX FUV and {\it Spitzer} and {\it Herschel} infrared data. We also use maps of dust properties, like the ones presented in \cite{rhc_aniano12}, based on the Draine \& Li dust model \citep{rhc_draine07} (DL07).

\subsection{Methods}

The native pixel size of the \cii\ maps after the reduction process was $2.7''$. We regrid the maps in order to define a new pixel size that is roughly the size of the \cii\ beam, i.e. $12''$. Among our data, the \cii\ maps have the lowest resolution, so we convolve all the other maps from the supplementary data to the \cii\ PSF. For this task, we use a library of convolution kernels for the cameras of the
{\it Spitzer}, {\it Herschel Space Observatory}, GALEX and ground-based telescopes constructed by \cite{rhc_aniano11}. After this step, we regrid the convolved maps to to be aligned with the \cii\ maps. 

For all the surface brightness and SFR surface density values we correct for inclination by a factor of cos~$i$.

\section{{\bf Results}} 

Our goal is to study if the \cii\ line can be used as a SFR tracer. To do this we compare the \cii\ line emission to four widely used SFR tracers: H$\alpha$, FUV, 24~$\mu$m and TIR emission. H$\alpha$ and FUV provide measures of star formation through the rate of production of ionizing photons and the photospheric emission from O and B stars, respectively. The 24~$\mu$m and TIR dust emission yields a measure of star formation via the reprocessing of light by dust in star-forming regions. Combination of these tracers is useful to account for the obscured (traced by 24~$\mu$m or TIR) and unobscured (traced by H$\alpha$ or FUV) contributions produced by star-forming regions. 

\begin{figure*}
\epsscale{1}
\plotone{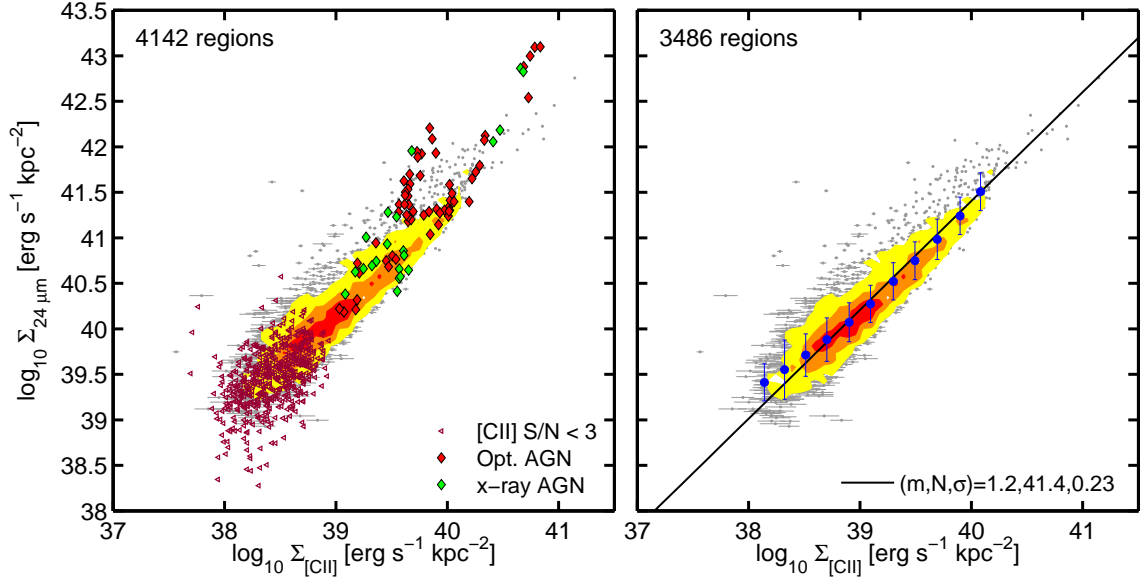}
\caption{ 24~$\mu$m surface brightness ($\Sigma_{24\mu\text{m}}$) versus [CII] 158~$\mu$m surface brightness ($\Sigma_{\rm [CII]}$) of 12'' (1 beam) size regions of 46 galaxies from the KINGFISH sample. Filled contours show the data density (similar to a Hess diagram), and enclose 90, 45 and 25\% of the data. ({\it Left}) Triangles correspond to \cii\ 3$\sigma$ upper limits of regions with $S/N<3$ ($\sim$11\% of the points). The diamonds represent the regions located within the central $\sim$0.5~kpc radius in 25 galaxies classified as AGN ($\sim$2\% of the points). Regions from the 19 galaxies optically classified as AGN are shown as red diamonds. Regions from the 6 galaxies that, based on the nuclear X-ray luminosity may indicate AGN activity, are shown as green diamonds. ({\it Right}) Same as left panel, but excluding points with [CII] $S/N < 3$, with emission associated with AGN.
The black line correspond to the best linear fit to the data (in the log--log space). The slope, the logarithmic value of $\Sigma_{24\mu\text{m}}$ at log$_{10}(\Sigma_{\rm [CII]}/[{\rm erg}~{\rm s}^{-1}~{\rm kpc}^{-2}]) = 40$ and the standard deviation dex of the fit are listed as (m,N,$\sigma$) on the bottom right corner. The blue dots show the running median and standard deviation in bins of $\Sigma_{\rm [CII]}$.\label{ciim24_all}}
\end{figure*}

\subsection{\cii\ -- 24~$\mu$m Correlation}

We start our analysis with the $\Sigma_{\rm [CII]} - \Sigma_{24~\mu m}$ correlation because we have 24~$\mu$m images available for all the galaxies in our sample. Figure \ref{ciim24_all} shows this correlation for 4,142 regions selected from KINGFISH galaxies. Three sigma upper limits for \cii\ emission are shown as dark red triangles. 

The 24~$\mu$m continuum emission is thought to be primarily produced by small hot dust grains in diffuse regions, transiently heated by the interstellar radiation field. One important contributor to the dust heating are young, hot stars \citep[e.g.][]{rhc_law11}; however, contribution from older stellar populations cannot be ignored. For example, in M~31, old stars contribute to the heating of the dust in star-forming regions in the disk  \citep{rhc_draine14} and dominate the heating in the bulge \citep{rhc_groves12}. In addition, non-stellar contribution to the radiation field by active nuclei can also be an important source of heating of dust grains \citep{rhc_dale06,rhc_deo09}. Given that our plan is to use the 24~$\mu$m to trace the young star formation that is reprocessed by dust, in the following section we describe how we account and correct for the 24~$\mu$m component that is not associated with star formation activity.

\begin{table*}
\begin{center}
\caption{Effect of cirrus correction on the $\text{[CII]} - 24\mu$m correlation\label{table1}}
\begin{tabular}{lccccc}
\\
\hline \hline
Description & Slope & Normalization\tablenotemark{a} & Scatter & $r_\text{corr}$ & Median $f_\text{cir}(-1\sigma,+1\sigma)$ \\
 & $m$ & N & [1$\sigma$ dex] &  &   \\
\hline
No Cirrus & 1.20 & 41.40 & 0.23 & 0.92 & 0 (0,0) \\
\hline
Scaled Cirrus \\ 
$U_\text{cir}=0.5 U_\text{min}$ & 1.21 & 41.32 & 0.24 & 0.89 & 0.18 (0.12,0.24) \\
..... $=0.75 U_\text{min}$ & 1.27 & 41.32 & 0.24 & 0.89 & 0.29 (0.19,0.39) \\
..... $=U_\text{min}$ & 1.30 & 41.26 & 0.27 & 0.87 & 0.39 (0.28,0.51) \\
\hline
Constant Cirrus \\
$U_\text{cir}=0.6$ & 1.30 & 41.41 & 0.24 & 0.90 & 0.15 (0.06,0.25) \\
..... $=0.8$ & 1.38 & 41.43 & 0.26 & 0.90 & 0.26 (0.11,0.42) \\
..... $=1.1$ & 1.60 & 41.50 & 0.33 & 0.87 & 0.43 (0.20,0.66) \\
\hline
\end{tabular}
\end{center}
\tablenotemark{a} The normalization N is the value of $\text{log}_{10}(\Sigma_{24\mu\text{m}})$ at $\text{log}_{10}(\Sigma_{\rm[CII]})=40$. Units are [erg~s$^{-1}$~kpc$^{-2}$]. \\
\end{table*}

\subsubsection{AGN Contribution}

X-ray photons produced by the AGN heat the surrounding dust and gas creating X-ray dominated regions (XDRs), \citep{rhc_maloney96}; as a result, XDRs can contribute to the total \cii\ and 24~$\mu$m dust emission. Even though by design of the sample the global luminosity of the KINGFISH galaxies is not dominated by AGN, the effect of the XDRs can be important in the central portions of galaxies. According to the nuclear spectral classification by \cite{rhc_moustakas10}, 19 of the 49 galaxies in our sample are optically classified as AGN and two other as mixed systems (SF/AGN). Another 6 galaxies that show no AGN signature in the optical have nuclear X-ray measured fluxes that may indicate AGN activity \citep{rhc_tajer05,rhc_grier11}. To study the effect of the AGN, we select the central region using circular apertures of $\sim0.5$~kpc radius. For galaxies further than $\sim$17~Mpc, the central 12" region is larger than 1~kpc; thus, for these cases we only mask the central 12" region.

The size of this aperture should be enough to enclose the emission arising from the XDRs powered by the AGN. The left panel on Figure~\ref{ciim24_all} shows the properties of the AGN-selected regions (color diamonds) and star-forming regions (grey dots) in the \cii\ -- 24~$\mu$m surface brightness plane. The AGN-selected regions are color coded according to the nucleus classification: optically selected AGN (red) and X-ray selected AGN (green). About half of the AGN-selected regions tend to show a 24~$\mu$m excess compared to \cii\ (or a \cii\ deficit compared to 24~$\mu$m); in the most extreme cases, the excess can be as high as a factor of $\sim$6 (e.g. regions from NGC~4736, see Appendix for individual correlations). It is likely that these higher 24~$\mu$m -- \cii\ ratios are caused by (1) the AGN contributing more to the dust continuum than to the \cii\ emission; (2) a reduction in the photoelectric heating efficiency due to the destruction of the small dust grains and PAHs by strong AGN radiation fields. Our goal is to use the 24~$\mu$m emission as a SFR tracer; thus, we remove from our sample the AGN-selected regions in order to avoid non star-forming contributions to the emission. 

Figure \ref{ciim24_all}, right panel, shows the $\Sigma_{24\mu\text{m}}$--$\Sigma_\text{[CII]}$ correlation after removing the \cii\ 3$\sigma$ upper limits and the AGN points. 
The black solid line represents the best linear fit to the remaining 3,486 points and the blue points show the median $\Sigma_{24\mu\text{m}}$ in bins of $\Sigma_{\text{[CII]}}$ with error bars indicating 1$\sigma$ scatter. The correlation is tight, with a 1$\sigma$ scatter around the fit of 0.23~dex and a slope of 1.20$\pm0.01$. This value is in excellent agreement with the 1.23 slope found by \cite{rhc_calzetti07} for the $\Sigma_{24\mu\text{m}}$ -- $\Sigma_\text{SFR}$ correlation; thus, we expect a nearly linear correlation between $\Sigma_{\rm [CII]}$ and $\Sigma_\text{SFR}$. We discuss more about the $\Sigma_{\rm [CII]}$ -- $\Sigma_\text{SFR}$ relationship in section 4.

\subsubsection{24~$\mu$m ``Cirrus'' Emission}

Emission at 24~$\mu$m can be used as a reliable obscured SFR tracer \citep{rhc_calzetti07, rhc_rieke09}; an important consideration, however, is to account for the 24~$\mu$m emission that is produced by dust heated by non star-forming sources (e.g., old stars). We will refer to this emission as 24~$\mu$m cirrus.

We estimate the intensity of the 24~$\mu$m cirrus following a similar procedure to the one described in \cite{rhc_leroy12}. The details of the cirrus calculation are presented in Appendix~C. The challenge in this method is to quantify the incident radiation field produced by non star-forming sources, i.e., $U_\text{cirrus}$. For our estimation of the 24~$\mu$m cirrus emission, we assume two distinct scenarios: (1) $U_{\rm cirrus}$ is constant across the galaxy, or (2) $U_{\rm cirrus}$ scales with $U_{\rm min}$ (in the DL07 model, $U_{\rm min}$ corresponds to the least interstellar radiation field heating the diffuse ISM). The resulting fraction of the 24~$\mu$m emission associated with cirrus ($f_\text{cir}$) and the effect of the cirrus correction on the $\Sigma_{24\mu\text{m}}$--$\Sigma_{\rm [CII]}$ correlation can be found in Table~\ref{table1}. In summary, 24~$\mu$m cirrus corrections based on a scaled version of $U_{\rm min}$ do not produce significant changes on the $\Sigma_{{\rm [CII]}}-\Sigma_{24~\mu{\rm m}}$ correlation, and the fraction of 24~$\mu$m cirrus emission is, on average, in the 18 to 39\% range (depending on the scaling factor assumed). We conclude that the results are robust to the choice of correction except when the correction is pushed to extreme values (e.g., $U_\text{cirrus}=1.1$). We know these extreme and likely fairly drastic cirrus assumptions are not representative of our local, $\sim$1~kpc neighborhood. Therefore, for the rest of the paper we choose to work with the $24~\mu$m cirrus subtraction that is based on the same assumption made by \cite{rhc_leroy12} for their sample of local galaxies, i.e. $U_\text{cirrus} = 0.5~U_\text{min}$. 

\subsection{\cii\ compared to other star formation tracers: H$\alpha$, FUV and 24~$\mu$m}

\begin{figure*}
\epsscale{1}
\plotone{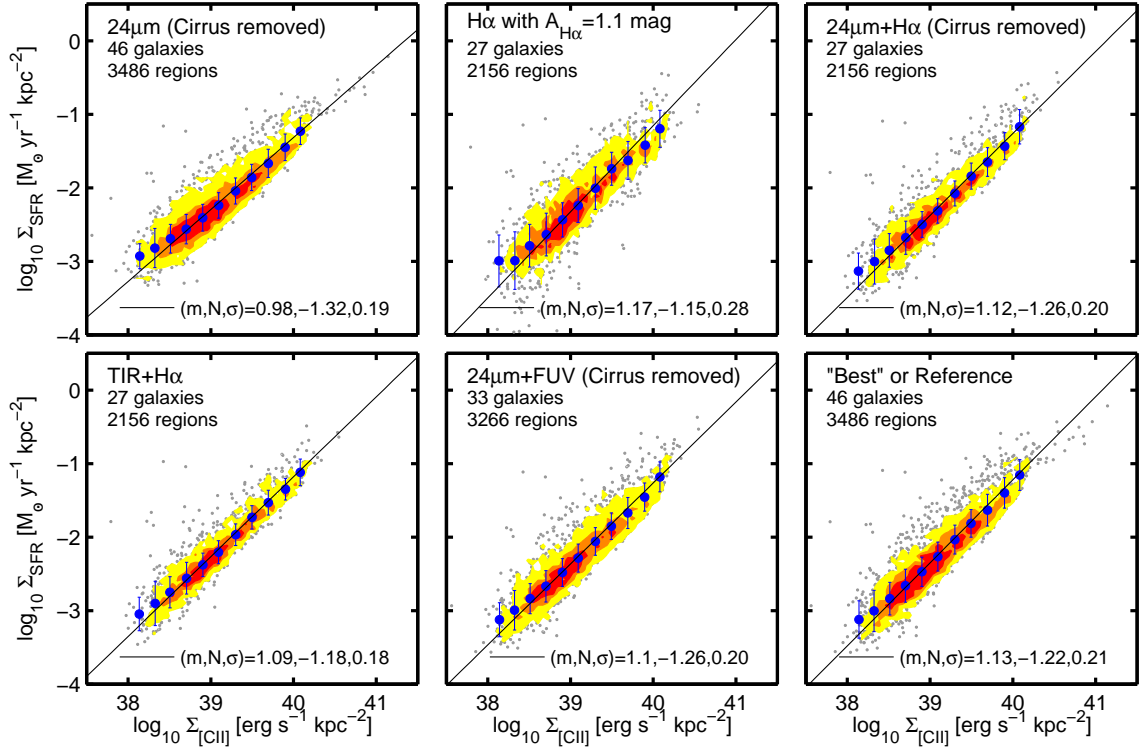}
\caption{Star formation rate surface density, $\Sigma_\text{SFR}$, versus [CII] 158~$\mu$m surface brightness, $\Sigma_{\rm[CII]}$, for 46 galaxies from the KINGFISH sample. Gray points correspond to 12'' (1 beam) size regions and filled contours show data density. The numbers on the bottom-right corner of each panel correspond to the fit parameters: slope ($m$), $y$-axis value at $x=40$ ($N$) and 1$\sigma$ standard deviation in dex ($\sigma$). Each panel shows the correlation for a different method of measuring $\Sigma_\text{SFR}$. ({\it Top-left}) We measure $\Sigma_\text{SFR}$ from 24~$\mu$m emission (after cirrus subtraction) following \cite{rhc_calzetti07} ({\it Top-middle}) For 27 galaxies with H$\alpha$ maps available, we measure $\Sigma_\text{SFR}$ from H$\alpha$ emission, corrected for internal extinction (1.1~mag), following \cite{rhc_calzetti07} calibration. ({\it Top-right}) We estimate $\Sigma_\text{SFR}$ based on the combination of 24~$\mu$m emission (cirrus subtracted) and H$\alpha$ emission for the 27 galaxies for which H$\alpha$ maps are available. We use \cite{rhc_calzetti07} calibration. ({\it Bottom-left}) We measure $\Sigma_\text{SFR}$ from the combination between TIR and H$\alpha$ emission following \cite{rhc_kennicutt09} calibration. ({\it Bottom-middle}) We measure $\Sigma_\text{SFR}$ as the combination of 24~$\mu$m emission (cirrus subtracted) and FUV emission for 33 galaxies with FUV maps available following \cite{rhc_leroy08} calibration. ({\it Bottom-right}) We show the ``Best'' or Reference $\Sigma_\text{SFR}$, that we measure from H$\alpha$+24~$\mu$m (when H$\alpha$ is available), FUV+24~$\mu$m (when H$\alpha$ is not available) and only 24~$\mu$m when neither H$\alpha$ nor FUV maps are available.
\label{ciisfr}}
\end{figure*}

On average, interstellar dust absorbs roughly half the starlight of typical spiral galaxies and re-emits it in the infrared; therefore, observations in the infrared are essential for deriving a complete inventory of star formation \citep{rhc_kennicutt12}. The best way to account for this dust-driven extinction is to combine unobscured star formation tracers, like H$\alpha$ or FUV emission, with the dust-reprocessed infrared continuum. In this section we explore the correlations between \cii\ and the SFR estimated from 24~$\mu$m and TIR emission and combinations of 24~$\mu$m and TIR with H$\alpha$ and FUV emission.

We measure the SFR and $\Sigma_{\rm SFR}$ based on: (1) H$\alpha$ emission using \cite{rhc_calzetti07} calculation, Equation (6) and applying a typical extinction correction of $A_{\text{H}\alpha}=1.1$~mag \citep{rhc_kennicutt83}. \cite{rhc_calzetti07} adopts the default $Starburst99$ IMF, i.e., a truncated Salpeter IMF with slope $1.3$ in the range $0.1-0.5~M_{\odot}$ and slope $2.3$ in the range $0.5-120~M_{\odot}$. (2) 24~$\mu$m emission using Equation~(8) and (9) in \cite{rhc_calzetti07}. (3) The combination of 24~$\mu$m and H$\alpha$ using \cite{rhc_calzetti07} calculation, Equation (7). (4) The combination of TIR luminosity and H$\alpha$ using \cite{rhc_kennicutt09} Equation (16) and a scaling coefficient of 0.0024 \citep[Table~4,][]{rhc_kennicutt09}. As \cite{rhc_calzetti07}, \cite{rhc_kennicutt09} also adopts the default $Starburst99$ IMF. We measure the TIR emission using 8, 24, 70 and 160~$\mu$m bands following Equation~(22) in \cite{rhc_draine07}. (5) The combination of  24~$\mu$m and FUV, using \cite{rhc_leroy08} Equation (D10) and (D11). These two calibrations were constructed from the FUV-based SFR calibration by \cite{rhc_salim07}. For all galaxies the 24~$\mu$m continuum emission is cirrus subtracted following \S3.1.2.

Figure~\ref{ciisfr} shows the correlations we find between $\Sigma_{\rm [CII]}$ and $\Sigma_\text{SFR}$. The first panel show the correlation for 46 galaxies for which we measure $\Sigma_\text{SFR}$ using only 24~$\mu$m emission. The slope of the correlation is nearly linear and the scatter is 0.19~dex. The next panel shows $\Sigma_\text{SFR}$ measured using H$\alpha$ for the 27 galaxies for which we have H$\alpha$ data available. The correlation is good ($r_{\rm corr}=0.83$), with the highest surface brightness points lying below the main trend probably because of increasing extinction in H$\alpha$. Given that H$\alpha$ is a tracer of recent star formation, with an age sensitivity of a few Myr \citep{rhc_mckee97}, the fact that we find a good correlation with \cii\ strengthens the case in favor of using this FIR cooling line as a SFR tracer. For the same 27 galaxies, the third and the fourth panel shows the combination between H$\alpha$ and 24~$\mu$m and TIR emission, respectively. Combining H$\alpha$ and IR continuum emission significantly reduces the scatter and corrects for the attenuation of H$\alpha$ emission at high SFR values. The fifth panel shows the $\Sigma_{\rm[CII]}$ -- $\Sigma_\text{SFR}$ correlation for 33 galaxies when using a combination of 24~$\mu$m and FUV emission to measure $\Sigma_\text{SFR}$.  The correlation is tight, with a 1~$\sigma$ scatter around the fit of 0.2~dex. The fit parameters are similar to those measured when the SFR is measured as a combination of H$\alpha$ and 24~$\mu$m emission.

Given that we do not have H$\alpha$ or FUV maps available for all the galaxies in our sample, unlike 24~$\mu$m, we define the $\Sigma_\text{SFR}$ we use from now on as ``our reference" $\Sigma_{\rm SFR}$ coming from the combination of: (1) 24~$\mu$m + H$\alpha$ (27 cases); (2) 24~$\mu$m + FUV if H$\alpha$ is not available (8 cases); (3) only 24~$\mu$m if neither H$\alpha$ nor FUV are available (11 cases). The correlation between the $\Sigma_{\rm[CII]}$ and ``our reference" $\Sigma_{\rm SFR}$ is shown in the last panel of Fig.~\ref{ciisfr}. The best linear fit to the data, as estimated by the OLS linear bisector method \citep{rhc_isobe90}, yields the following relationship:

\begin{multline} \label{eq:correlation_uncorr}
\Sigma_{\rm SFR}~({\rm M}_{\odot}~{\rm yr}^{-1}~{\rm kpc}^{-2}) = 3.79\times10^{-47} \\
\times (\Sigma_{\rm[CII]}~[{\rm erg~s^{-1}~kpc^{-2}}])^{1.13}.
\end{multline}

\noindent The scatter of the correlation is 0.21~dex. 

In order to derive a calibration for the SFR based on [CII] luminosities, we convert the [CII] luminosity surface densities into [CII] luminosities, and then we fit the data using the OLS linear bisector method. The resulting [CII]--based SFR calibration is:

\begin{multline} \label{eq:correlation_lum}
\rm SFR~({\rm M}_{\odot}~{\rm yr}^{-1}) = 2.29\times10^{-43} \\
\times (L_{\rm[CII]}~[{\rm erg~s^{-1}}])^{1.03}.
\end{multline}

\noindent The scatter of the correlation is 0.21~dex. Due to the distance effect introduced by the conversion to luminosities, the calibration in equation~(\ref{eq:correlation_lum}) has a slope closer to unity, but similar scatter. Recall that ``luminosity-luminosity" relations implicitly have distance squared in both axes ($\propto D^{2}$), while in ``surface density-surface density'' correlations, the quantities in both axes depend upon the ratio between luminosity ($\propto D^{2}$) and area ($\propto D^{2}$), so surface densities are independent of distance.

Before applying these calibrations, it is important to understand their reliability and limits of applicability. For instance, the luminosity calibration is subject to the caveats mentioned in the Introduction regarding the ``\cii-deficit", which implies that galaxies with similar IR luminosity can show variations of a factor of 10x or more in their \cii\ luminosity \citep{rhc_stacey10}. In order to explore the reliability of the \cii--SFR calibration, in the next section we study in detail the nature of the scatter in the \cii--SFR correlation and we also apply these calibrations to other samples of extragalactic objects observed in \cii\ emission.

\section{Analysis}

In the previous section we found a tight correlation between $\Sigma_{\rm [CII]}$ and $\Sigma_\text{SFR}$. In this section we try to understand the origin of this correlation and the reason why some galaxies or regions within galaxies deviate from this quasi-linear relationship.

Variations of the \cii\ luminosity compared to the IR continuum were first observed by ISO. Low \cii\ to FIR ratios are found for global measurements of normal star-forming galaxies with warm dust temperatures ($F_{\nu}(60~\mu\text{m})/F_{\nu}(100~\mu\text{m})\gtrsim0.8$) {\citep{rhc_malhotra97,rhc_malhotra01,rhc_brauher08} and luminous and ultraluminous infrared galaxies (U/LIRGs; $L_\text{IR}>10^{11-12}~L_{\odot}$) {\citep{rhc_luhman98,rhc_luhman03}. This is important in the context of our study because the FIR luminosity is commonly used a SFR tracer in U/LIRGs. Therefore, any variation in the \cii\ to FIR ratio will imply a difference between the SFR measured using \cii\ and FIR emission.

With {\it Herschel}, low \cii\ to FIR ratios are observed for U/LIRGS as a function of increasing dust temperature, compactness of the source \citep{rhc_diaz-santos13} and FIR luminosity to molecular gas mass ($M_{\rm H_{2}}$) ratio \citep{rhc_gracia-carpio11}. In addition, {\it Herschel} allowed for the first time to resolve the regions that exhibit low \cii\ to FIR ratios in nearby galaxies. \cite{rhc_croxall12} observe in NGC~4559 and NGC~1097 a drop in the \cii\ to FIR ratio for regions with warm dust temperatures ($\nu f_{\nu}(70~\mu\text{m})/\nu f_{\nu}(100~\mu\text{m})\gtrsim0.95$) and intense radiation fields. They conclude that the most plausible scenario to account for the \cii\ deficit is the charging of the dust grains caused by the high radiation fields. \cite{rhc_beirao12} find a similar trend when comparing the circumnuclear ring and extranuclear regions of NGC~1097. 

Similar to the observed variations in the \cii\ to FIR ratios, \cite{rhc_delooze14} report variations in the \cii\ to SFR correlation observed in galaxies from the DGS sample. Compared to a standard SFR measured as a combination of FUV and 24~$\mu$m emission, they find systematically lower values of \cii-based SFRs as a function of increasing dust temperature, \oi~63~$\mu$m/\cii+\oi~63~$\mu$m ratio, and decreasing metallicity. Thus, for warm, low metallicity regions \cite{rhc_delooze14} conclude that \oi~63~$\mu$m is a more reliable SFR tracer than \cii. For additional discussion on the connection between metal abundance and the dispersion of the \cii--SFR correlation, see \cite{rhc_delooze14} .

In order to evaluate the reliability of a SFR calibration based on the \cii\ line, it is key to understand the scatter in the correlation. The goal is to identify the variables that drive the deviations from the fit (equations~\ref{eq:correlation_uncorr} and \ref{eq:correlation_lum}), and use this information to reduce the scatter and establish the limits of applicability of the calibration. In the next two sections we discuss how local regions within a galaxy and galaxies as a whole deviate from the fit. We study the deviations as a function of a set of parameters that characterize the ISM properties and can be derived directly from the dust continuum SEDs and spectra. These parameters are: (1) IR color, $\nu f_{\nu}(70~\mu\text{m})/\nu f_{\nu}(160~\mu\text{m}$); (2) oxygen abundance ($12+log({\rm O/H})$); (3) fraction of the IR luminosity radiated from regions with high radiation fields, $f(L_\text{IR}; U>100$); (4) dust-weighted mean starlight intensity, $\langle U \rangle$, and (5) PAH abundance, $q_\text{PAH}$. These last three parameters are derived from the DL07 dust model (see Appendix~B for details).

\begin{figure*}
\epsscale{1}
\plotone{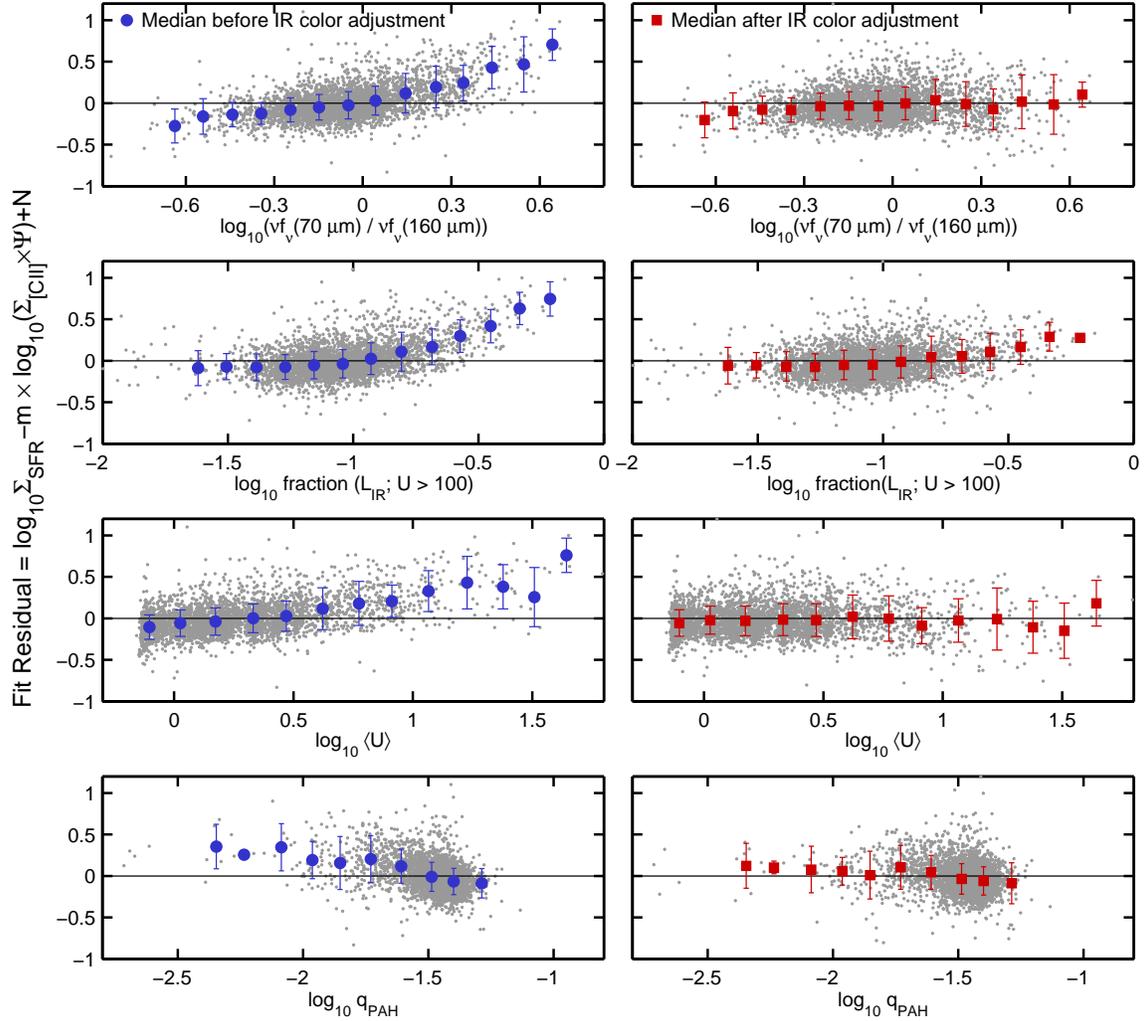}
\caption{Fit residual of the $\Sigma_{\rm[CII]} - \Sigma_\text{SFR}$ correlation as a function of: IR color ($\nu f_{\nu}(70)/\nu f_{\nu}(160)$), fraction of the IR luminosity radiated from regions with high radiation field (fraction(L$_\text{IR};U>100$)), median radiation field ($\langle U \rangle$) and PAH abundance ($q_\text{PAH}$). The left and the right panel show the fit residual before and after applying the IR color adjustment derived in \S4.2.1, respectively. Gray points represent 12'' size regions of the 46 galaxies in the sample. Filled blue dots (left panel), red squares (right panel) and the corresponding vertical bars correspond to the median and the 1~$\sigma$ standard deviation in dex of the binned distribution of points.\label{dev_local}}
\end{figure*}

\begin{table}
\begin{center}
\caption{IR Color Adjustment Coefficients\label{IRcolor_correction}}
\begin{tabular}{lccc}
\hline \hline \\
\multicolumn{4}{c}{${\rm log}_{10}(\Sigma_{\rm SFR}) = m \times {\rm log}_{10}(\Sigma_{\rm [CII]}\times\Psi(\gamma))+N$} \\ \\
\hline
IR Color ($\gamma$)  & Threshold ($\gamma_{t}$) & $\alpha$ \\ \hline
$\nu f_{\nu}(70~\mu\text{m})/\nu f_{\nu}(160~\mu\text{m}$) & 1.24 & 0.94\\
$\nu f_{\nu}(70~\mu\text{m})/\nu f_{\nu}(100~\mu\text{m}$) & 0.80 & 1.57\\
\hline \\
\multicolumn{4}{c}{${\rm log}_{10}({\rm SFR}) = m \times {\rm log}_{10}(L_{\rm [CII]}\times\Psi(\gamma))+N$} \\ \\
\hline
IR Color ($\gamma$)  & Threshold ($\gamma_{t}$) & $\alpha$ \\ \hline
$\nu f_{\nu}(70~\mu\text{m})/\nu f_{\nu}(160~\mu\text{m}$) & 1.12 & 1.20\\
$\nu f_{\nu}(70~\mu\text{m})/\nu f_{\nu}(100~\mu\text{m}$) & 0.80 & 1.90\\
\hline \\
\multicolumn{4}{l}{Note: Coefficients $m$ and $N$ can be found in Table~\ref{calibration_coeff}}
\end{tabular}
\end{center}
\end{table}

\subsection{Local Variations}

\begin{figure*}
\epsscale{0.8}
\plotone{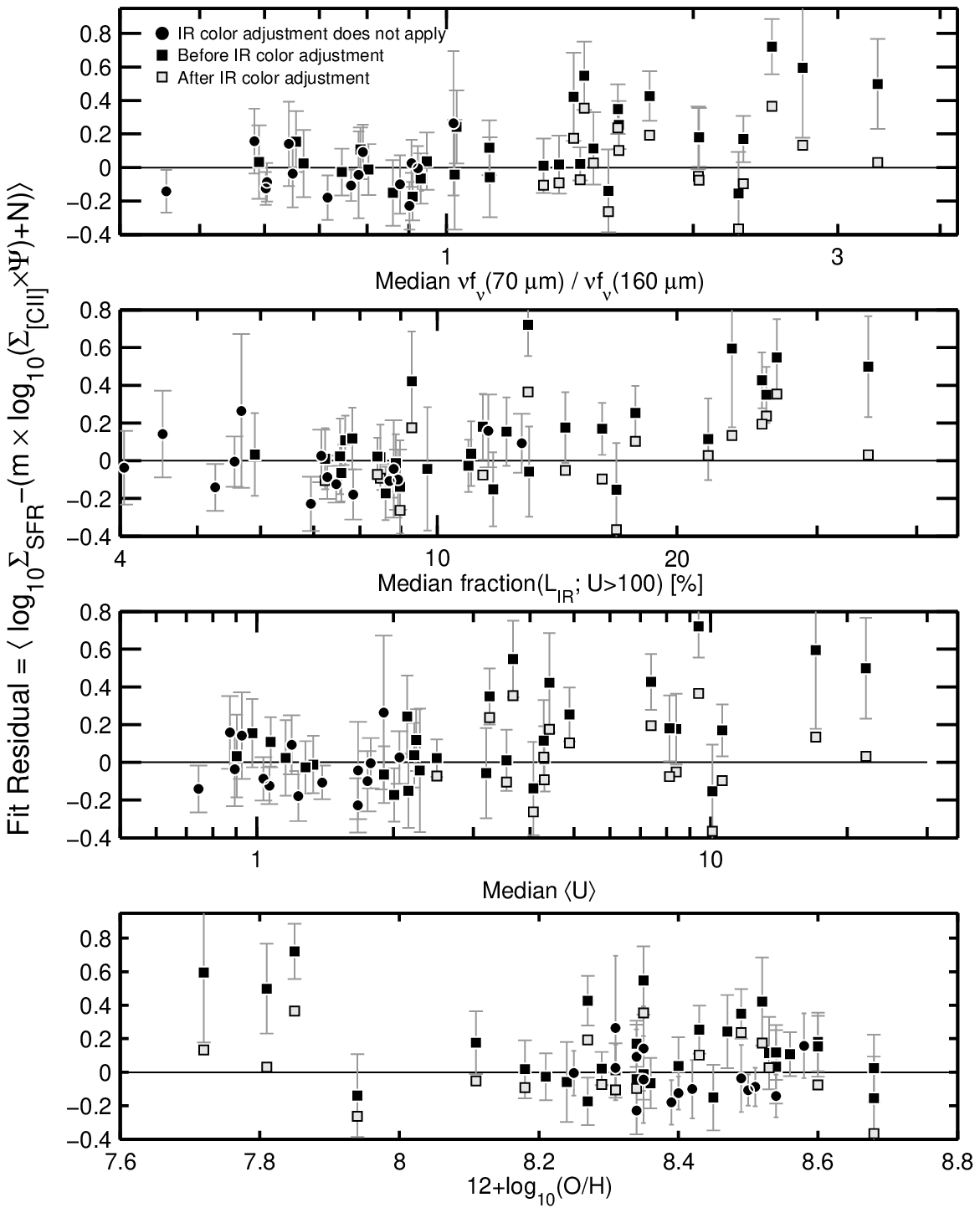}
\caption{Median of the fit residual for each of the 46 galaxies in our sample versus median IR color ($\nu f_{\nu}(70~\mu\text{m})/\nu f_{\nu}(160~\mu\text{m}$), median radiation field ($\langle U \rangle$), median fraction of the luminosity coming from regions with $U>100$ and oxygen abundance ($12+\text{log(O/H)}$). The vertical bars represent 1-$\sigma$ standard deviation around the median. Black squares and open squares show the median fit residual before and after applying the IR color adjustment. Black circles show the median fit residual for galaxies that, on average, are not warm enough to apply the IR color adjustment.\label{dev_global}}
\end{figure*}

The high spatial resolution provided by {\it Herschel} allow us to study the scatter on $\sim$kiloparsec scales. The benefit of studying local variations in a large sample of galaxies with varying global properties is that allow us to identify the more likely of the possible scenarios behind the scatter. Figure~\ref{dev_local} summarizes the results of our fit residual analysis for all the $\sim12$\arcsec  regions of the 46 galaxies in our sample. The top-left panel shows the fit residual as a function of IR color, which is an observable quantity and can be used as a proxy for the dust temperature. About 20\% of the regions have IR colors warmer than $\nu f_{\nu}(70~\mu\text{m})/\nu f_{\nu}(160)~\mu\text{m}\gtrsim1.3$; the fit residual for these regions systematically increases as a function of IR color, with a median deviation from the fit as high as a factor of $\sim3$ for the warmest regions (i.e., warm dust regions systematically exhibit higher $\Sigma_{\rm SFR}$ (or SFR) than expected from their \cii\, meaning they are underluminous in \cii). This corresponds to a particular dust color temperature, $T_\text{dust}$. We estimate $T_\text{dust}$ using a modified blackbody model with emissivity spectral index $\beta=1.5$ and the 70 to 160~$\mu$m flux ratio. Most of the regions have dust color temperatures in the range $T_\text{dust}\sim20-32$~K, and we find that regions start to systematically deviate from the fit for dust color temperatures higher than $T_\text{dust}\gtrsim 29$~K. 

A similar trend is observed as a function of the model-derived variables. The remaining left-panels of Figure~\ref{dev_local} show that for a dust-weighted mean starlight intensity of $\langle U \rangle \gtrsim 6$, a fraction of the IR luminosity produced in high radiation field regions $f({L_\text{IR};U>100}) \gtrsim 25\%$, or a PAH abundance $q_{PAH}\lesssim2\%$,  points tend to deviate from the fit in a similar fashion as for IR color. It is important to note that $\sim$99\%, $\sim$85\% and $\sim$73\% of the regions with $\langle U \rangle \gtrsim 6$, $f({L_\text{IR};U>100}) \gtrsim 25\%$ and $q_{PAH}\lesssim2\%$ have IR colors $\gtrsim1.3$, respectively; this shows that regions that systematically deviate from the fit in the different panels are essentially the same. We also observed a similar systematic deviation as a function of the $\nu f_{\nu}(70~\mu\text{m})/\nu f_{\nu}(100~\mu\text{m})$ IR color, starting at the threshold value of $\nu f_{\nu}(70~\mu\text{m})/\nu f_{\nu}(100~\mu\text{m})\approx0.8$. 

We also analyze the local variations as a function of dust attenuation, which we measure as the fraction of FUV and optical emission absorbed and reprocessed by dust versus the escaping FUV emission as the ratio of the 24~$\mu$m to the FUV intensity, $I_{24\mu\rm{m}}/I_{\rm{FUV}}$. The details of this analysis can be found in Appendix~D. Around the $I_{24\mu\rm{m}}/I_{\rm{FUV}}$ ratio of $\sim10$, we observe a systematic increase of the fit residuals as a function of decreasing dust attenuation. Low metallicity regions from Holmberg~II and IC~2574 also show high fit residuals at low $I_{24\mu\rm{m}}/I_{\rm{FUV}}$ ratios, and we discuss the implications of these results in \S4.2.1.


Before analyzing the possible physical reasons behind the fit residual increase described in the previous paragraph, we need to discard that these deviations are not introduced by systematic disagreements between the different SFR tracers combined to produce our reference $\Sigma_{\rm SFR}$ (\S3.2). In order to explore this possibility, we first construct two alternative versions of the reference $\Sigma_{\rm SFR}$: one that is solely based on the combination of $24~\mu$m and H$\alpha$, and one that is based on the combination of $24~\mu$m and FUV only. For both cases, we observe a similar systematic increase of the fit residuals as a function of $\langle U \rangle$ as the one shown in Figure~\ref{dev_local}. Therefore, we discard that the way we construct our reference $\Sigma_{\rm SFR}$ has any implication on the observed trend of the fit residuals. 

Next, we try to identify the physical basis for the systematic increase of the fit residuals. As we mentioned before, at least $\sim$85\% of the regions with warm IR colors ($\gtrsim1.3$) show evidence of high radiation fields. These regions also show lower PAH abundances, which is expected given that it appears that PAHs are destroyed in HII regions \citep{rhc_povich07}. One of the possible reasons for the systematic increase of the residual towards positive values is the decrease of the photoelectric heating efficiency due to the positive charging of the dust grains and the decrease in the gas heating rates due the lower abundance of PAHs. What this implies for the purpose of measuring the $\Sigma_{\rm SFR}$ (or SFR) based on $\Sigma_{\rm[CII]}$ (or $L_{\rm[CII]}$) alone is that a simple calibration that ignores dust color temperature will underestimate the amount of star formation activity in these regions. 

\subsubsection{IR Color Adjustment}

Motivated by the observed systematic increase of the fit residual after a given IR color threshold, we derived an IR color adjustment that accounts for the underestimation of $\Sigma_\text{SFR}$ (or SFR). We choose an adjustment based on an IR color, over a model-dependent variable adjustment (e.g., $\langle U \rangle$), because the IR color is an observable, free of assumptions and easier to measure. 
For a given IR~color $\gamma(\lambda_{1},\lambda_{2})=\nu f_{\nu}(\lambda_{1}~\mu{\rm m})/\nu f_{\nu}(\lambda_{2}~\mu{\rm m})$, we define the IR color factor $\Psi$ as:

\begin{equation} \label{eq:IRcorrection}
\Psi(\gamma) = \begin{cases} 1 & \mbox{if } \gamma \mbox{~$\textless~\gamma_{\rm t}$} \\ (\gamma/\gamma_{\rm t})^{\alpha} & \mbox{if } \gamma \mbox{~$\geq\gamma_{\rm t}$.} \end{cases}
\end{equation}

\noindent The IR~color threshold, $\gamma_{\rm t}$, and the power law exponent, $\alpha$, are derived in order to minimize the logarithmic residuals between the observed and the \cii-based $\Sigma_{\rm SFR}$ (or SFR). Values of $\gamma_{\rm t}$ and $\alpha$ for the IR~colors $\gamma(70,100)$ and  $\gamma(70,160)$ are listed in Table~\ref{IRcolor_correction}. This proposed IR~color based adjustment $\Psi$ represents a simple and straightforward attempt to account for the systematic increment of the fit residual of regions that show warm colors/high radiation fields. Based on the IR~color factor $\Psi$, the adjusted \cii\ surface brightness (or luminosity) can now we written as $\Sigma_{\rm[CII]}\times\Psi$ (or $L_{\rm [CII]}\times\Psi$).

The effects of the IR color adjustment on the correlation residuals are shown in the right panels of Figure~\ref{dev_local}. The red squares show the adjusted median residuals as a function of IR color, $\langle U \rangle$, $f({L_\text{IR};U>100})$ and $q_\text{PAH}$. By design, the IR color adjustment removes the trend of increasing residuals with IR color for warmer regions. This is true even for the last bin in IR color where the difference between the adjusted and the unadjusted data is a factor of $\sim$3. As the second and third left panels show, applying the IR color adjustment also helps to remove the trend of increasing residuals with $\langle U \rangle$ and $f({L_\text{IR};U>100})$. This is a consequence of the large overlap between regions with warm colors and high radiation field signatures. In the case of the PAH abundance, the fourth panel show that the IR color adjustment helps to reduce the increasing residuals as a function of $q_\text{PAH}$. This is expected given the correlation between IR color, $\langle U \rangle$ and $q_\text{PAH}$.

\begin{figure}
\epsscale{1.2}
\plotone{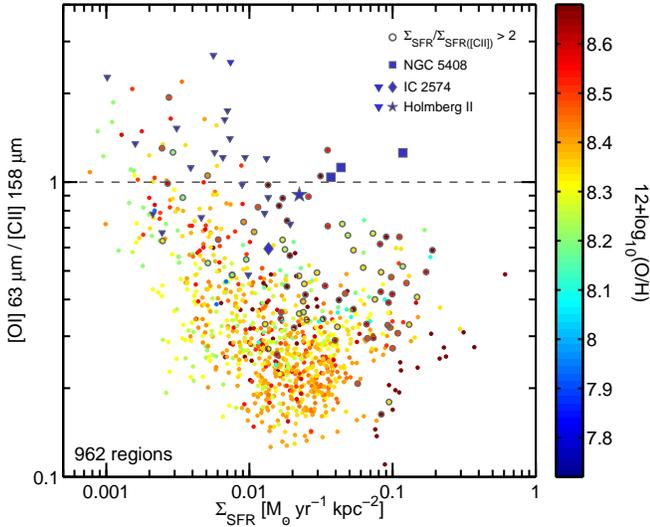}
\caption{Ratio between the [OI]~63~$\mu$m and [CII]~158~$\mu$m emission for regions selected from the KINGFISH sample as a function of the SFR surface density ($\Sigma_\text{SFR}$) measured as the combination of 24~$\mu$m, H$\alpha$ and FUV emission (see \S3.2). The color scale shows the global metallicity value of the galaxy from \cite{rhc_moustakas10}. The circles with grey border show regions where this $\Sigma_\text{SFR}$ value is greater than the [CII]-based $\Sigma_\text{SFR}$ by more than a factor of two. The panel also shows regions selected from three low metallicity systems: Holmberg~II (star), IC~2574 (diamond) and NGC~5408 (square). The triangles pointing down correspond to 3-$\sigma$ upper limits due to non-detections in [OI]~63~$\mu$m emission. These low metallicity regions also tend to show [CII]-based $\Sigma_\text{SFR}$ values that are lower than $\Sigma_\text{SFR}$ by a factor of two or more (see Figure~\ref{dev_global}, panel 4).\label{oicii}}
\end{figure}

\subsection{Galaxy-to-Galaxy Variations}

In this section we continue the study of the scatter of the correlation, but this time in terms of the galaxy-to-galaxy variations. Treating each galaxy as a single point is an oversimplification of the underlying physics, but it is useful because frequently only a global measurement of the \cii\ flux can be obtained.

In order to analyze how much each galaxy deviates from the fit as a whole, we fix the slope to that determined for the $\Sigma_\text{[CII]}$--$\Sigma_\text{SFR}$ correlation (i.e. 1.13$\pm$0.01, eq.~\ref{eq:correlation_uncorr}) and then we calculate the median residual for each galaxy. Figure~\ref{dev_global} shows the median fit residual as a function of IR color ($\nu f_{\nu}(70)/\nu f_{\nu}(160)$), dust-weighted mean starlight intensity $\langle U \rangle$, fraction of the IR luminosity radiated from regions with high radiation fields, $f(L_\text{IR}; U>100)$ and the global measurements of oxygen abundance (12+log$_{10}$(O/H)). Each point represents one galaxy. Those with mean IR~colors below the IR~color adjustment threshold (i.e. $\gamma_{t}(70,160)=1.24$) are shown as filled circles. The rest of the galaxies, for which the IR color adjustment applies, are shown as filled and open squares for unadjusted and adjusted mean residual values, respectively. 

The first panel of Figure~\ref{dev_global} shows the median residuals as a function of IR color.  Starting around $\nu f_{\nu}(70)/\nu f_{\nu}(160)\sim1.2$, we observe a trend of increasing residuals with warmer IR color, similar to what we find in the analysis for the resolved regions (\S4.1). The effect of the IR color adjustment on the residuals is clearly shown by the growing separation between the unadjusted and adjusted mean residual values as a function of IR color. The second and the third panels show the residual as a function of two radiation field strength related parameters: $\langle U \rangle$ and $f(L_\text{IR}; U>100)$. We find that systems with $\langle U \rangle\gtrsim3$ or $f(L_\text{IR}; U>100)\gtrsim20\%$ start to show increasing deviations from the fit. This is similar to what we observe in the IR color panel and expected given the close connection between the radiation field strength and the IR color. As we previously showed in the analysis of the resolved regions, it is clear from these two panels how the IR color adjustment helps to reduce the absolute scatter in the correlation. In terms of the standard deviation of the median residuals, applying the IR color adjustment reduces the galaxy-to-galaxy scatter from 0.22~dex to 0.16~dex. 

\subsubsection{Low Metallicity Galaxies}

The KINGFISH sample contains only a handful of low metallicity systems, which despite their paucity show some potentially interesting trends. The bottom panel in Figure~\ref{dev_global} shows that low metallicities galaxies tend to deviate from the fit. What this implies for the three of them -- Holmberg~II, IC~2574, and NGC~5408 --, is that the \cii-based calibration yields a SFR that is at least a factor of $\sim4$ lower than the SFR measured using the standard tracers available (i.e. FUV+24~$\mu$m for Holmberg~II and IC~2574, and 24~$\mu$m for NGC~5408). These are diffuse galaxies with no well-defined bulge or nucleus, so we discard potential AGN contamination in the SFR measurement \citep{rhc_moustakas10}. In the case of the irregular dwarf galaxy IC~2574, the \cii\ data covers only the HI supergiant shell located in the northeast part of the galaxy discovered by \cite{rhc_walter98}. This region contains a central stellar cluster about 11~Myr old and recent star formation activity located in the rim of the expanding shell \citep{rhc_stewart00}. This far-UV intense environment heats the surrounding dust grains, resulting in high dust temperatures. In fact, this region from IC~2574, together with Holmberg~II and NGC~5408, correspond to the three warmest systems in our sample as measured by their IR colors: for an emissivity index of $\beta=1.5$, their dust color temperatures are in the T$_{\rm dust}=30-34$~K range. In addition, these galaxies have \cii\ to total infrared (TIR) luminosity ratios in the $0.35-0.6\%$ range. This is slightly higher than the 0.3\% measured for the Milky Way \citep{rhc_wright91}. Compared to other low metallicity systems, these ratios are lower than those found for NGC~4214 \citep[$0.5-0.8\%$,][]{rhc_cormier10} and the Large~Magellanic~Cloud \citep[0.9\%,][]{rhc_rubin09}, and higher than those measured in 30~Doradus \citep[0.12\%,][]{rhc_poglitsch95} and Haro~11 \citep[0.1\%,][]{rhc_cormier12}.

There are multiple factors that can play a role in the high $\Sigma_{\rm SFR}/\Sigma_{\rm SFR([CII])}$ ratios observed in low metallicity regions \citep[e.g.,][]{rhc_delooze14}. If we assume that it is correct to use the same method to measure $\Sigma_{\rm SFR}$ in both high and low metallicity regions, then the high residuals can only be explained by variations in the heating and \cii-cooling as a function of metallicity. On one hand, in low-metallicity environments low PAH abundances, low dust-to-gas ratios, and low FUV extinction reduce the gas heating efficiency and heating rate. On the other, FUV photons produced by O and B stars travel farther from their point of origin, thereby producing a low diffuse FUV flux in much of the ISM that may keep dust grains mostly electrically neutral and maintain a high gas heating efficiency \citep{rhc_israel11}. Note, however, that we measure the highest dust temperatures, implying high dust-weighted mean starlight intensities ($\langle U \rangle>10$), in our low metallicity sample.

A possible explanation of the deviations observed in low metallicity regions is that we are missing a non-negligible fraction of the neutral gas cooling, coming out in FIR lines other than \cii\ (e.g., \oi~$63~\mu$m). In order to explore this scenario, we measure the ratio of the \oi~$63~\mu$m to \cii\ emission as a function of the SFR surface density. Figure~\ref{oicii} shows the line ratio for 962 regions in the KINGFISH sample for which we have \oi~$63~\mu$m and \cii\ detections with $S/N>3$. At low $\Sigma_{\rm SFR}$ values, regions tend to have higher \oi~63~$\mu$m to \cii\ line ratios compared to the rest of the sample. To investigate this, we have done data simulations in which we apply different $S/N$ cuts, and we find that the absence of regions with small line ratios at low $\Sigma_{\rm SFR}$ is not real, but most likely a bias introduced by the selection of the $S/N=3$ cut. From Figure~\ref{oicii} we conclude that the \oi~$63~\mu$m line emission is not the dominant cooling channel for the KINGFISH metal-rich regions. We also include in the plot the three low metallicity systems in our sample: NGC~5408 (blue) and 3-$\sigma$ upper limits for Holmberg~II (red) and IC~2574 (magenta). Together with these low metallicity regions, the other regions in our sample that show $\Sigma_{\rm SFR}$ to $\Sigma_{\rm SFR([CII])}$ ratios greater than two are shown as circles with grey border. It can be seen that these regions tend to show higher \oi~$63~\mu$m to [CII] line ratios than the rest of the sample. In the case of NGC~5408, the cooling of the neutral gas is approximately equally split between the [CII] and \oi~$63~\mu$m transitions. This result is similar to the \cii\ to \oi~63$\mu$m ratios of 1.1 and 1.2 measured by \cite{rhc_hunter01} for the irregular galaxies NGC~1569 and IC~4662, respectively. For Holmberg~II and IC~2574, the upper limits do not rule out the possibility of a non-negligible contribution to the cooling of the neutral gas via \oi~$63~\mu$m. What this implies for low metallicity regions, is that a purely \cii-based SFR calibration would underestimate the total SFR value; for NGC~5408, this would account for at least half of the factor of $\sim4$ deviation. 

Can \oi~$63~\mu$m um emission be used to provide a better SFR estimate? \cite{rhc_delooze14} find that a \cii-based SFR calibration systematically underestimates the SFR as a function of increasing \oi~$63~\mu$m/\cii+\oi~$63~\mu$m ratio. For regions with sub-solar metallicity, they conclude that \oi~$63~\mu$m is a more reliable SFR tracer than \cii\ emission. Over the same KINGFISH sample we study in \cii, the $\Sigma_{\rm [OI]}-\Sigma_{\rm SFR}$ correlation has a slope of 1.2 and a scatter of 0.25~dex. This is less linear and has higher scatter than what we measure in the $\Sigma_{\rm [CII]}-\Sigma_{\rm SFR}$ correlation. It is possible that the sample of dwarf galaxies in \cite{rhc_delooze14} is biased towards starbursting systems, where stronger radiation fields arising from dense PDRs favor \oi\ as a more reliable star formation tracer than \cii. If we combine the \cii\ and \oi~$63~\mu$m emission in our sample, we find that the slope and the scatter of the correlation between $\Sigma_{\rm [CII]+[OI]}-\Sigma_{\rm SFR}$ are similar to that of the $\Sigma_{\rm [CII]}-\Sigma_{\rm SFR}$ correlation.

\begin{figure*}
\epsscale{1}
\plotone{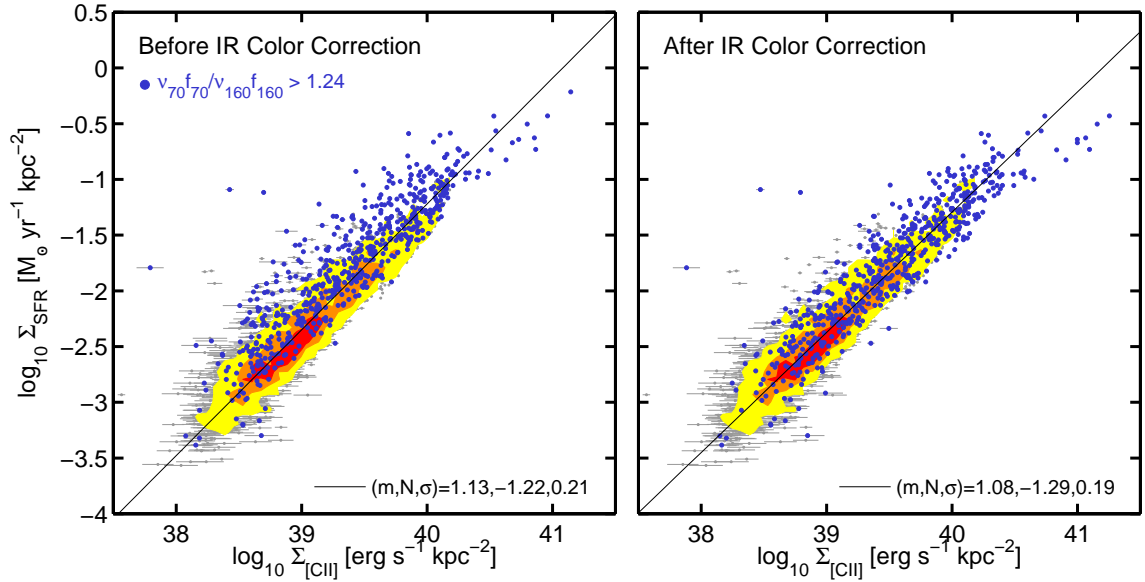}
\caption{ {\it (Left)} Star formation rate surface density, $\Sigma_\text{SFR}$, as a function of [CII]~158~$\mu$m surface brightness, $\Sigma_{\rm[CII]}$, for 46 galaxies from the KINGFISH sample. $\Sigma_\text{SFR}$ is estimated from H$\alpha$+24~$\mu$m (when H$\alpha$ is available), FUV+24~$\mu$m (when H$\alpha$ is not available) and only 24~$\mu$m when neither H$\alpha$ nor FUV maps are available. Gray points correspond to 12\arcsec~(1 beam) size regions and contours show data density. The numbers on the bottom-right corner of each panel correspond to the fit parameters: slope ($m$), $y$-axis value at $x=40$ ($N$) and 1$\sigma$ standard deviation ($\sigma$). {\it (Right)} Same as left panel, but for this case we adjusted the $\Sigma_{\rm[CII]}$ using the IR color adjustment described in equation~\ref{eq:IRcorrection}.
\label{ciisfr_best}}
\end{figure*}

Perhaps more importantly, at lower metallicities we expect lower dust optical depth, resulting in regions that are more transparent to the FUV radiation than their metal-rich counterparts because of low dust-to-gas ratios. Under these conditions, the fraction of FUV photons that escapes the system without interactions with the dust is larger than at normal metallicity. In dense PDRs -- the main source of \oi\ emission -- the FUV leakage would play against the production of both lines, \cii\ and \oi, for which interaction of FUV photons with the dust is required for heating. We speculate, however, that the lower density material that emits preferentially in \cii\ is more affected by the escape of FUV photons, giving rise to higher \oi~63~$\mu$m/\cii\ ratios in these sources. This would help to explain the high \oi~63~$\mu$m/\cii\ ratios observed in the low-metallicity regions compared to the rest of the sample.

A higher fraction of escaping FUV photons at low metallicity would also imply that the IR emission would not be able to account for the total value of the SFR. As an example, \cite{rhc_calzetti07} find that using 24~$\mu$m emission can underestimate the SFR of low metallicity systems by a factor of $\sim2-4$. In order to explore this scenario we measure the FUV to \cii\ and FUV to TIR ratios, which can be used as a rough measure of the amount of extinction at ultraviolet wavelengths \citep{rhc_dale07}. We find that the mean FUV to \cii\ ratio (and the corresponding 68\% range in parenthesis) for the low metallicity regions from Holmberg~II and IC~2574 is $281(134-588)$, whereas the remaining 31 galaxies with FUV data available have a median FUV to \cii\ ratio of $32(13-77)$. We observe a similar trend for the FUV to TIR ratios. Thus our results favor the scenario described above, where a larger fraction of escaping FUV photons in low metallicity regions would mostly explain the low SFRs based on \cii\ emission. 

\subsubsection{Dependence of the correlation fit parameters with distance and inclination}

The galaxies in our sample span a range in distance ($2.8-26.5$~Mpc) and inclination ($\lesssim75^{\circ}$) that might have potential consequences on the individual $\Sigma_\text{[CII]}$--$\Sigma_\text{SFR}$ correlation properties. To test this, we look for trends between the correlation fit parameters and the distance or inclination of the source. We fit the $\Sigma_\text{[CII]}$--$\Sigma_\text{SFR}$ correlation for each individual galaxy using the OLS bisector method (we leave out of this analysis four galaxies that have fewer than 5 pointings with S/N$>3$). We do not find any clear trend between the fit parameters (i.e., slope and normalization) with distance or inclination. The Pearson correlation coefficient for the four correlations made as the combination of the slope/normalization versus the distance/inclination is less than $\sim0.2$ for all the cases. Therefore we conclude that the calibrations presented here are robust to spatial resolution and inclination. 

\subsection{The [CII] -- star formation rate correlation before and after the IR color adjustment}

As we show in \S4.1 and \S4.2, it is possible to use an IR color based adjustment to reduce the scatter of both the $\Sigma_{\rm SFR}$--$\Sigma_{\rm[CII]}$ and SFR--$L_\text{[CII]}$ correlations. The consequences of applying these adjustments are shown in Fig.~\ref{ciisfr_best}. The left panel shows the $\Sigma_{\rm[CII]}-\Sigma_{\rm SFR}$ correlation before applying the IR color adjustment. The regions that are warm enough to be modified by the adjustment ($\nu f_{\nu}(70)/\nu f_{\nu}(160)\geq1.25$) are shown as blue dots. The black solid line correspond to the best fit to the data (eq.~\ref{eq:correlation_uncorr}). The right panel on Fig.~\ref{ciisfr_best} shows the correlation after applying the IR color adjustment. The net effect is that the asymmetry of the scatter cloud decreases, reducing the dispersion of the correlation to 0.19~dex. The OLS bisector fit to the IR color corrected correlation yields a slope that is slightly closer to linear (1.08). 

Table~\ref{calibration_coeff} summarizes the best-fit parameters we measure for the $\Sigma_{\rm SFR}$--$\Sigma_{\rm[CII]}$ and SFR--$L_\text{[CII]}$ correlations. We also list the 1-$\sigma$ dispersion around the fit and the correlation coefficient $r_{\rm corr}$. We include the calibration coefficients for four different cases: (1) the correlation including the 24~$\mu$m normal cirrus subtraction. These correspond to the coefficients in equation~(\ref{eq:correlation_uncorr}) and equation~(\ref{eq:correlation_lum}); (2) same as (1), but after applying the IR color adjustment from equation~(\ref{eq:IRcorrection}); (3) the correlation after applying the strong 24~$\mu$m cirrus subtraction (which assumes $U_{\rm cirrus}=U_{\rm min}$); (4) if we fit the correlation using a fixed slope of 1. For regions or galaxies in the surface brightness range of the KINGFISH sample ($10^{39}\lesssim \Sigma_{{\rm[CII]}} \lesssim10^{40.5}$~[erg~s$^{-1}$~kpc$^{-2}$]), the resulting SFRs using the calibration parameters from the first two cases are very similar. However, the calibration coefficients from case 1 (Eq.~\ref{eq:correlation_uncorr} and Eq.~\ref{eq:correlation_lum}) seem better suited to measure SFRs in higher surface brightness objects (e.g., LIRGs) as can be seen in the next section.

As a summary, the \cii-based calibration can be written as:

\begin{equation} \label{eq:SFcalibration}
{\rm log}_{10}(\Sigma_{\rm SFR}) = m \times ({\rm log}_{10}(\Sigma_{\rm [CII]}\times\Psi(\gamma))-40)+N
\end{equation}

\noindent where the calibration coefficients $m$ and $N$ are listed in Table~\ref{calibration_coeff}. We recommend using the calibration coefficients from Equation~\ref{eq:correlation_uncorr}. If there is IR color information available, we also recommend to use Table~\ref{IRcolor_correction} to measure the IR color factor $\Psi$ and apply the IR color adjustment to the \cii\ emission.

\begin{figure}
\epsscale{1}
\plotone{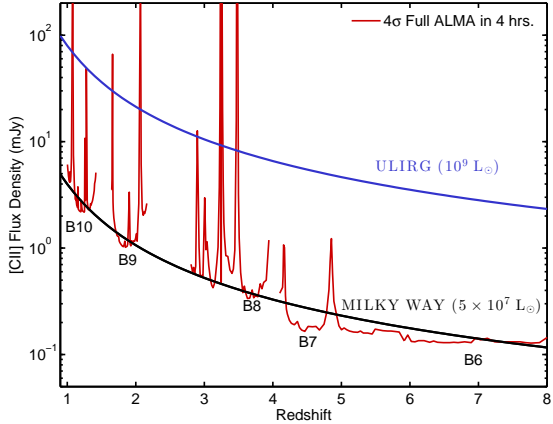}
\caption{ALMA's ability to detect Milky Way type \citep[black line at 5$\times10^7~{L_{\odot}}$;][]{rhc_wright91,rhc_bennett94} and ultraluminous infrared type galaxies (blue line at $10^{9}$~L$_{\odot}$) in the [CII]~158~$\mu$m transition after 4 hours of time integration as a function of redshift. The red line represents ALMA's sensitivity estimated using the ALMA sensitivity calculator (assuming a linewidth of 300~km/s). From Band~10 (B10) to Band~6 (B6), galaxies can be detected by ALMA in the [CII] transition in the redshift range of $z\sim1.2-7$.\label{cii_alma}}
\end{figure}

\begin{figure*}
\epsscale{1}
\plotone{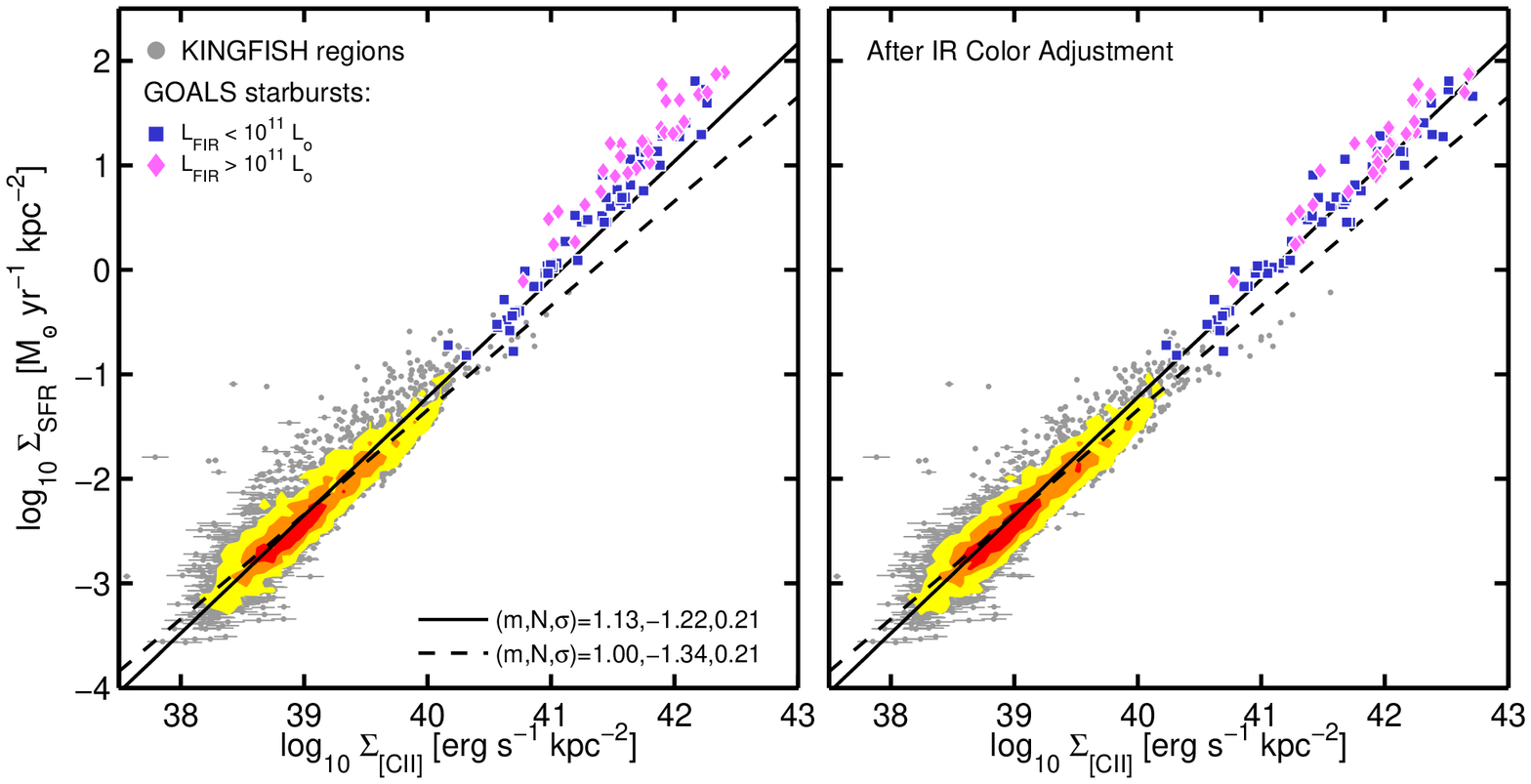}
\caption{{\it (Left)} Star formation rate surface density versus [CII] surface brightness for resolved regions from the KINGFISH sample (grey points and color density contours) and global measurements from a subset of pure starburst systems from the GOALS sample. These were selected using their 6.2~$\mu$m PAH EW. The blue squares and the magenta diamonds show the GOALS galaxies that have IR luminosities lower and higher than $10^{11}$~L$_{\odot}$, respectively. The solid and the dashed line correspond to Equation~\ref{eq:correlation_uncorr} and the fit with a fixed slope of 1, respectively. The numbers on the bottom-right corner correspond to the fit parameters: slope ($m$), $y$-axis value at $x=40$ ($N$) and 1$\sigma$ standard deviation in dex ($\sigma$). These are the same parameters as in equation~\ref{eq:correlation_uncorr}. {\it (Right)} Same as left panel, but for this case we adjusted the $\Sigma_{\rm[CII]}$ in both samples using the IR color adjustment described in equation~\ref{eq:IRcorrection}. The solid and dashed line are the same as in the left panel. \label{ciisfr_density}}
\end{figure*}

\section{Comparison with Models and Observations}

\subsection{Comparison to other extragalactic \cii\ samples}

Comparing our results with previous analyses done for samples of galaxies observed in \cii\ by {\it ISO} and {\it Herschel} can provide useful information about the reliability and limits of applicability of our \cii-based SFR calibration. This is particularly true for very luminous galaxies, which are poorly represented in the KINGFISH sample. This comparison can be done in surface density or luminosity space. Surface density comparisons are the most interesting from a physical standpoint, since they are distance independent. Moreover, surface density is more directly related to parameters like radiation field intensity, which are thought to dominate the physical processes in PDRs. Luminosity, on the other hand, is frequently the only measurement directly available for very distant galaxies, and it is thus also interesting to investigate. For instance, Figure~\ref{cii_alma} shows ALMA's ability to detect, in the \cii\ transition, galaxies like the Milky Way and ULIRGs as a function of redshift. For an integration time of 4~hours, the full ALMA can potentially detect in the \cii\ transition Milky Way type galaxies at redshift intervals that span the range from $z\sim1.2$ to 7.

\begin{table*}
\begin{center}
\caption{[CII]--SFR calibration coefficients and uncertainties\label{calibration_coeff}}
\begin{tabular}{lcccc}
\hline \hline \\
\multicolumn{5}{c}{${\rm log_{10}}(\Sigma_{\rm SFR}/[{\rm M}_{\odot}~{\rm yr}^{-1}~{\rm kpc}^{-2}])=m\times({\rm log_{10}} (\Sigma_{\rm [CII]}/[{\rm erg}~{\rm s}^{-1}~{\rm kpc}^{-2}])-40)+N$} \\ \\
\hline
Description & Slope & Normalization\tablenotemark{a} & Scatter & $r_\text{corr}$ \\
 & $m$ & $N$ & [1$\sigma$ dex] &   \\ \hline
Normal 24~$\mu$m Cirrus (Eq.~\ref{eq:correlation_uncorr}) & $1.13\pm0.01$ & $-1.22\pm0.01$ & 0.21 & 0.92\\
IR Color Adjusted & $1.08\pm0.01$ & $-1.29\pm0.01$ & 0.19 & 0.93\\
Strong 24~$\mu$m Cirrus & $1.18\pm0.01$ & $-1.25\pm0.02$ & 0.23 & 0.90\\
Fixed Slope & $1.00$ & $-1.34\pm0.01$ & 0.21 & 0.92\\
\hline \\
\multicolumn{5}{c}{${\rm log_{10}}({\rm SFR}/[{\rm M}_{\odot}~{\rm yr}^{-1}])=m\times({\rm log_{10}} ({\rm L_{\rm [CII]}}/[{\rm erg}~{\rm s}^{-1}])-40)+N$} \\ \\
\hline
Description & Slope & Normalization\tablenotemark{b} & Scatter & $r_\text{corr}$  \\
 & $m$ & $N$ & [1$\sigma$ dex] &   \\ \hline
Normal 24~$\mu$m Cirrus (Eq.~\ref{eq:correlation_lum}) & $1.03\pm0.01$ & $-1.28\pm0.01$ & 0.22 & 0.94\\
IR Color Adjusted & $0.98\pm0.01$ & $-1.40\pm0.01$ & 0.20 & 0.96\\
Strong 24~$\mu$m Cirrus & $1.03\pm0.01$ & $-1.02\pm0.01$ & 0.24 & 0.92\\
Fixed Slope & $1.00$ & $-1.32\pm0.01$ & 0.22 & 0.94\\
\hline
\end{tabular}
\end{center}
\tablenotemark{a} The normalization $N$ is the value of $\text{log}_{10}(\Sigma_{\text{SFR}}/[{\rm M}_{\odot}~{\rm yr}^{-1}~{\rm kpc}^{-2}])$ at $\text{log}_{10}(\Sigma_{\rm[CII]}/ [{\rm erg}~{\rm s}^{-1}~{\rm kpc}^{-2}])=40$.\\
\tablenotemark{b} The normalization $N$ is the value of $\text{log}_{10}({\rm SFR}/[{\rm M}_{\odot}~{\rm yr}^{-1}])$ at $\text{log}_{10}(L_{\rm[CII]}/ [{\rm erg}~{\rm s}^{-1}])=40$.\\
\end{table*}

First, we do the comparison in the surface density space. For this purpose, we use galaxies from the Great Observatories All-sky LIRG Survey \citep[GOALS]{rhc_armus09}. A brief description of the selection criteria and properties of the GOALS systems included in this analysis is available in Appendix~D. Figure \ref{ciisfr_density} shows the $\Sigma_{\rm SFR}-\Sigma_{\rm[CII]}$ correlation for the KINGFISH regions and the GOALS galaxies; the latter are divided in pure starburst with $L_{\rm FIR}<10^{11}L_{\odot}$ and non-AGN U/LIRGs. The left and right panels show the correlation before and after applying the IR color adjustment, respectively. The solid line in both panels represents equation~(\ref{eq:correlation_uncorr}). The dashed line correspond to the best linear fit to the KINGFISH regions (Table~\ref{calibration_coeff}). For the data with no IR color adjustment applied, the GOALS pure starburst galaxies with $L_{\rm FIR}<10^{11}L_{\odot}$ and $\Sigma_{\rm[CII]}\lesssim10^{41.5}$~[erg~s$^{-1}$~kpc$^{-2}$] agree well with equation~(\ref{eq:correlation_uncorr}). At greater luminosity surface densities, these systems start to systematically deviate from the fit, although they still lie roughly within a factor of two from equation~(\ref{eq:correlation_uncorr}). On the other hand, all the non-AGN U/LIRGs lie above the fit, following a trend parallel to the KINGFISH data. If we fit the GOALS pure starburst with $L_{\rm FIR}<10^{11}L_{\odot}$ and non-AGN U/LIRGs using the slope from equation~(\ref{eq:correlation_uncorr}), we find that the fit normalization is, on average, a factor 1.5 and 3 higher, respectively. When compared to the linear fit, GOALS starbursts galaxies with $\Sigma_{\rm[CII]}\gtrsim10^{41}$~[erg~s$^{-1}$~kpc$^{-2}$] show deviations in the $\sim5-10$ range. As we show in \S4.1, part of these deviations are associated to the IR color of the region. Therefore, we apply the IR adjustment derived for the KINGFISH sample to the regions and galaxies in both samples. The right panel in Figure \ref{ciisfr_density} shows the IR color adjusted correlation, where the good agreement between the KINGFISH and the GOALS samples with Equation~(\ref{eq:correlation_uncorr}) across nearly 5 orders of magnitude in $\Sigma_{\rm[CII]}$ and $\Sigma_{\rm SFR}$ is evident. On the other hand, the linear \cii-based calibration -- even after applying the IR color adjustment -- continue to underestimate the reference $\Sigma_{\rm SFR}$ value.

Next, we do the comparison in luminosity space. In addition to the galaxies from the GOALS sample, we include in the analysis galaxies used by \cite{rhc_boselli02}, \cite{rhc_delooze11} and \cite{rhc_sargsyan12} to derive their \cii-based SFR calibrations. We also include non-AGN, non-merger LIRGs from Weiner et al. (in prep.). The general properties of these additional samples of galaxies are described in Appendix~D. In order to compare our KINGFISH results to the other samples, we re-derive their published SFRs and TIR luminosities, so we can ensure uniformity and comparability to our work. 

The results of the comparison between the KINGFISH galaxies and the samples described above are shown in Fig.~\ref{ciisfr_compilation}. The two panels show the ratio of the \cii-based SFR calibration (Equation~\ref{eq:correlation_lum}) to the SFR measured using standard tracers as a function of TIR luminosity. Each black point corresponds to the mean SFR of a KINGFISH galaxy, and the vertical bars represent the 1-$\sigma$ standard deviation around the mean value. The TIR luminosity of the KINGFISH galaxies was measured over the area covered by the \cii\ observations and not the entire system. The left panel shows the galaxies before applying the IR color adjustment. We see that the largest deviations occur at both TIR luminosity ends. At $L_{\rm TIR}\lesssim10^{9}~L_{\odot}$, 6 out of the 8 systems that show deviations larger than a factor of $\sim3$ have metallicites below $12+{\rm log}_{10}{\rm (O/H)}\lesssim8.2$: Holmberg~II and IC~2574 from the KINGFISH sample; NGC~625, NGC~1569 and NGC~1156 from \cite{rhc_delooze11} sample; IC~4662 from \cite{rhc_boselli02} sample. The remaining two galaxies, NGC~4698 and NGC~4429 from \cite{rhc_boselli02} sample, do not have metallicity measurements available. At the high TIR luminosity end, the SFR measured using the \cii--SFR calibration underestimates the SFR(TIR) value for almost all the LIRGs. The non-AGN LIRGs show deviations that can be as high as a factor of $\sim10$. If we fit these systems using the slope from Equation~\ref{eq:correlation_lum}, we find that the fit normalization is, on average, a factor of $4.4\pm1.9$ higher. Similarly, \cite{rhc_delooze14} find that in the $L_{\rm [CII]}-\rm{SFR}$ plane, ULIRGs tend to be offset from starburst and AGNs by a factor between 3 and $\sim$10. These deviations are a direct consequence of the significant scatter in the \cii\ to TIR ratio observed in U/LIRGs \citep{rhc_malhotra97,rhc_malhotra01,rhc_brauher08,rhc_diaz-santos13}. It is important to mention, however, that monochromatic IR-based SFR tracers (e.g., 24~$\mu$m) agree much better with the SFR inferred from \cii\ through our calibration; but integrated indicators such as TIR are usually considered better measures of the SFR in U/LIRGs.

\begin{figure*}
\epsscale{1}
\plotone{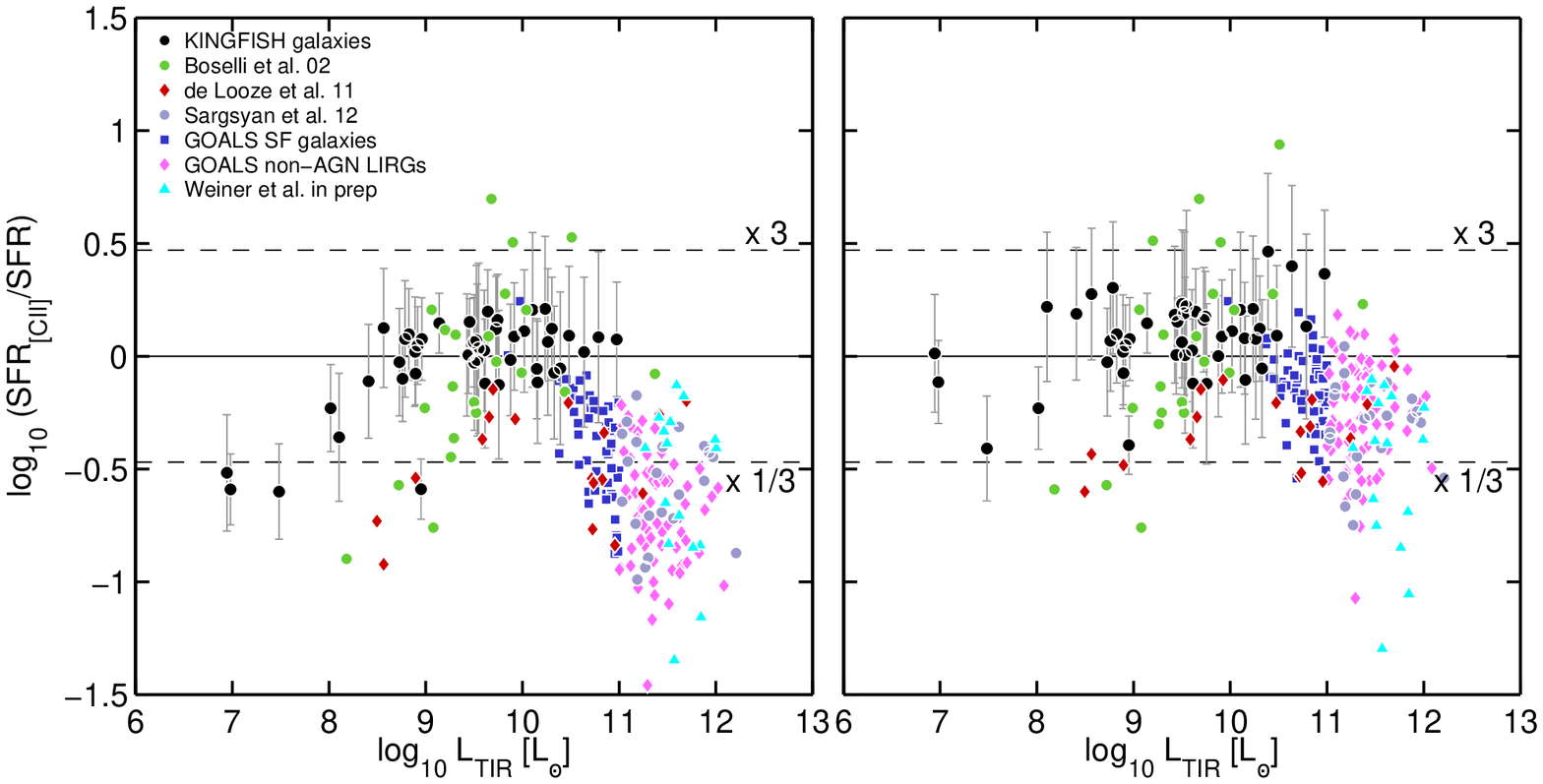}
\caption{Ratio of the SFR measured using this work [CII] calibration (Eq.~\ref{eq:correlation_lum}) to the SFR measured based on tracers other than [CII]. We treat each KINGFISH galaxy as an individual point (black circle) by taking the median value of the SFR and the sum of the TIR luminosity of all the regions in the galaxy that were observed in [CII]. The vertical bar represents the 1-$\sigma$ standard deviation around the SFR median. The other galaxy global measurements correspond to \citealt{rhc_boselli02} (green circles), \citealt{rhc_delooze11} (red diamonds), \citealt{rhc_sargsyan12} (purple circles), \cite{rhc_diaz-santos13} (blue squares and magenta diamonds) and Weiner et al. in prep (cyan circles). The left and right panels show the data before and after applying the IR color adjustment. \label{ciisfr_compilation}}
\end{figure*}

The right panel shows the result of applying the IR color adjustment derived in our study. As shown in \S4.1, the IR color adjustment helps to reduce the discrepancy for the low-metallicity KINGFISH galaxies; this is also true for the low metallicity systems from the other samples. For the LIRGs, the IR color adjustment proves to have an important effect by reducing the difference between the SFR(\cii) and SFR(TIR) values to less than a factor of two for more than half of the systems. The IR color adjustment is especially effective with the $L_{\rm IR}<10^{11}L_{\odot}$ pure starbursts from the GOALS sample. The LIRGs of the \cite{rhc_sargsyan12} sample, however, are not corrected enough to bring them into good agreement with the calibration. It is possible that the IR color of these galaxies is redder than the color of the regions from which most of the \cii\ arises, or maybe a the color adjustment -- although adequate for the KINGFISH sample -- is too mild for many of these systems. If we fit all the LIRGs using the same slope as equation~\ref{eq:correlation_lum}, we find that the fit normalization is a factor of 1.9 higher. That is, the LIRGs can be placed on a relation approximately parallel to our calibration for normal galaxies, but displaced toward higher SFR per \cii\ emission. Note, however, that there is a large scatter of the U/LIRGs around this offset.

In summary, for samples of normal, star-forming galaxies and non-AGN LIRGs, the SFR measured using our \cii--SFR calibration -- after applying IR color adjustment -- agrees within a factor of $\sim3$ with the SFR from standard SFR tracers (H$\alpha$, FUV, 24~$\mu$m and FIR) for at least $\sim80 \%$ of the systems. The remaining galaxies can exhibit deviations as high as factor of $\sim$10, showing the limitations of using the \cii\ luminosities to measure SFRs in IR luminous and ultra-luminous systems. This is not the case for the $\Sigma_{\rm [CII]}-\Sigma_{\rm SFR}$ correlation, where the KINGFISH regions and the GOALS starbursts agree within a factor of $\sim2$ with the calibration (Equation~\ref{eq:correlation_uncorr}) over 5 orders of magnitude in $\Sigma_{\rm [CII]}$. This calibration is more physically motivated than the luminosity one, because surface densities are connected to the radiation field strength, which has a strong influence on the drivers of the \cii\ emission process. In the luminosity case, the same \cii\ luminosity can be attained by a combination of high/low [CII] surface brightness over a small/large emitting area. The physical conditions in these two scenarios are significantly different, resulting in large deviations from the calibration for a given fixed \cii\ luminosity. 

\subsection{Comparison with Models: Starburst99}

In regions where the cooling is dominated by \cii\ emission, we expect the FUV heating and the \cii\ cooling to be closely related through the photoelectric effect in PAHs and small dust grains. In this section we explore this connection using a simple model based on the {\it Starburst99} code \citep{rhc_leitherer99}. 

The first step in our calculation is to use {\it Starburst99} to model the luminosity of a stellar population for a constant SFR over 100~Myr. For the calculations we adopt the default evolutionary tracks, and assume a stellar population with solar metallicity. For the stellar initial mass function (IMF), we adopt the {\it starburst99} default. The output of the model is the spectrum of a stellar population as a function of the duration of the star formation episode ($t_{\rm d}$). From this spectrum, we measure the FUV luminosity by integrating over the FUV range of energy which dominates the grain photoelectric heating, i.e. $6 < E_{\gamma} < 13.6$~eV. Then, we scale the FUV luminosity by assuming a heating efficiency, $\epsilon_{\rm h}$, to estimate the amount of heating of the gas. $\epsilon_{\rm h}$ is the product of two factors: (1) the photoelectric heating efficiency of the dust ($\epsilon_\text{ph}$), that is mainly set by the ratio of the photoionization rate over the recombination rate of electrons with neutral grains/PAHs \citep{rhc_hollenback99,rhc_weingartner01}. Typical values for $\epsilon_\text{ph}$ are in the $0.1 - 3\%$ range and, (2) the fraction of the FUV photons in the $6 < E_{\gamma} < 13.6$~eV range that interact with dust and result in gas heating of the neutral and molecular ISM. We assume that the cooling of the gas is dominated by the \cii\ transition, so for a given SFR, we connect the \cii\ and FUV luminosities via $L_{\rm[CII]} \sim \epsilon_{h} \times L_\text{FUV}(t_{\rm d})$ (as we show in Fig.~\ref{oicii}, this is a valid assumption for metal-rich regions).  

Figure~\ref{ciisfr_sb99} shows the model results plotted on top of the $\Sigma_{\rm[CII]}$ -- $\Sigma_\text{SFR}$ correlation.  Each line represents a model output for a combination of $t_{\rm d}$ and $\epsilon_\text{h}$. Dashed lines correspond to a population with a constant star formation rate and $t_{\rm d}=2$~Myr; the solid lines correspond to $t_{\rm d}\geq20$~Myr (because we are assuming a constant star formation rate scenario, the variation of the FUV luminosity in the $20-100$~Myr range is less than $\sim$25\%). As expected, for a given $t_{\rm d}$ and $\Sigma_\text{SFR}$, as $\epsilon_\text{h}$ increases, so does the predicted $\Sigma_{\rm[CII]}$. Note that there is a degeneracy in the model between $t_{\rm d}$ and $\epsilon_\text{h}$. For instance, the model output for the combination of input parameters $t_{\rm d}=2$~Myr and $\epsilon_\text{h}=3\%$ is very similar to the model output when assuming $t_{\rm d}\geq20$~Myr and $\epsilon_\text{h}=1\%$. 

Figure~\ref{ciisfr_sb99} also shows individual star-forming regions (no IR color adjustment applied) selected from 46 KINGFISH galaxies. The data density contours represent the ensemble of all these regions. The bulk of the data can be explained as arising from regions that have been actively star-forming for $t_{\rm d}>20$~Myr, with $\epsilon_\text{h}$ in the 1--3\% range. This suggests that the scatter is mainly driven by changes in $\epsilon_\text{h}$. On the other hand, to describe the bulk of the data assuming $t_{\rm d}=2$~Myr requires an unusually high heating efficiency of $\epsilon_\text{h} > 3\%$. 

The model suggests two possible explanations to describe the regions that deviate the most from the main trend and tend to lie in the upper side of the scatter cloud: (1) an early stage of the star formation episode ($t_{\rm d}=2$~Myr) and $\epsilon_\text{h}$ in the 1--3\% range, or (2)  standard star formation duration of $t_{\rm d}\geq20$~Myr and low $\epsilon_\text{h}$ ($\lesssim1$\%). The second scenario is widely applicable because it does not require fine tuning of the ages, and we know that at high UV fields small grains become positively charged or are destroyed, thus reducing $\epsilon_\text{h}$.  

Finally, Figure~\ref{ciisfr_sb99} highlights four individual cases. One of them, NGC~4254, is a spiral galaxy chosen to illustrate a system that follows the main trend. The other three are:

{\it NGC~2146}: this system has the highest total IR luminosity in our sample, $L_\text{TIR} = 10^{11}$~L$_{\odot}$ and can be classified as a LIRG. The regions from NGC~2146 have [CII] luminosity surface densities in the range $10^{40}\lesssim\Sigma_{\rm[CII]}\lesssim10^{41.3}$~[erg~s$^{-1}$~kpc$^{-2}$]. The majority of them follow the main trend and can be described by heating efficiencies in the $1-3\%$ range and $t_{\rm d}\geq20$~Myr.

{\it NGC~2798}: about half of the regions of this galaxy show a higher ratio of $\Sigma_{\rm SFR}$ to $\Sigma_{\rm[CII]}$ compared to the rest of the data. NGC~2798 is a barred spiral that is part of an interacting pair with NGC~2799. There is evidence from UV spectra for a recent burst of star formation \citep{rhc_joseph86}. This agrees well with the calculations for a duration of the star formation of about $t_{\rm d}=2$~Myr and $\epsilon_\text{h}$ in the range 1--3\%.

{\it Galaxies with $12+{\rm log(O/H)}<8.1$}: here we select regions from four low metallicity galaxies that have oxygen abundances $12+{\rm log(O/H)}<8.1$.  These systems are: Holmberg~II, IC~2574, NGC~2915, and NGC~5408 \citep{rhc_moustakas10}. As shown in Fig.~\ref{dev_global}, about half of the low metallicity regions deviate from the fit showing higher $\Sigma_{\rm SFR}/\Sigma_{\rm [CII]}$ ratios than the rest of the points. In these systems two scenarios are possible: (1) the regions are young and have been forming stars at a continuous rate for only $t_{\rm d}=2$~Myr, and the heating efficiency is in the $\epsilon_{h}\simeq1-3\%$ range; (2) the regions have been forming stars at a continuous rate for $t_{\rm d}\geq20$~Myr and the heating efficiency of the medium is lower than $\epsilon_{h}\sim1\%$. As we concluded in \S4.2.2, the deviations observed in low metallicity regions are most likely described by the latter scenario, where a reduction in the heating efficiency is expected due to reduced trapping of FUV photons by the dust.

\begin{figure}
\epsscale{1}
\plotone{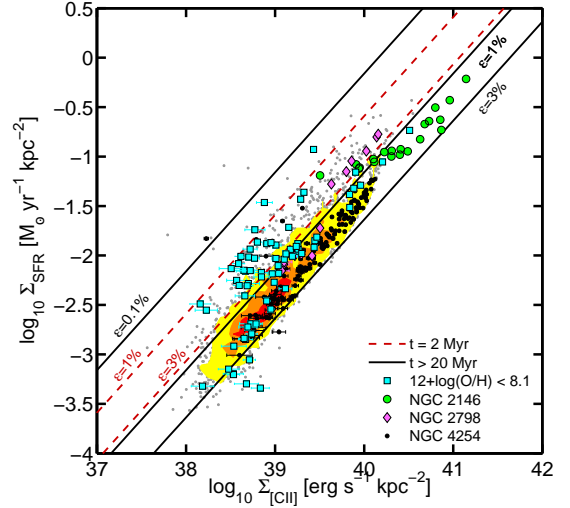}
\caption{Star formation rate surface density versus [CII] surface density (no IR color adjustment applied). The gray points and the corresponding data density contours show the star-forming regions selected from 48 galaxies of the KINGFISH sample. We also highlight individual galaxy cases: NGC~2146 (green circles), NGC~2798 (magenta diamonds), NGC~4254 (black circles) and regions from low metallicity galaxies with 12+log(O/H)$\lesssim8.1$ (cyan squares). The diagonal lines represent the model output. We use {\it Starburst99} to model the FUV luminosity of a stellar population for a given SFR value and a constant star formation scenario. We then convert the FUV emission into a [CII] surface brightness by assuming a heating efficiency of, $\epsilon_\text{h}$. Thus, each diagonal line is the combination of a $t_{\rm d}$~Myr old population with constant star formation rate and $\epsilon_\text{h}$. Dashed lines correspond to $t_{\rm d}=2$~Myr;  the solid line correspond to $t_{\rm d}\geq20$~Myr.\label{ciisfr_sb99}}
\end{figure}

\section{Summary and Conclusions}

We study the \cii~158~$\mu$m line emission and its potential to be used as a star formation rate tracer using a set of nearby galaxies drawn from the {\it Herschel} KINGFISH sample. The \cii\ surface brightness, $\Sigma_{\rm [CII]}$, can be used as a robust $\Sigma_{\rm SFR}$ tracer in normal, star-forming galaxies in the absence of strong AGNs. In this work we present a calibration for that relation (Equation~\ref{eq:correlation_uncorr}) that is based on 3,486 regions selected from 46 nearby galaxies. The uncertainty associated with the calibration is $\pm0.21$~dex. One of the main sources of scatter are regions with warm IR colors. We derive a set of adjustments based on the IR color factor $\Psi$ (Equation~\ref{eq:IRcorrection} and Table~\ref{IRcolor_correction}) that helps to reduce the scatter among the warmer regions. Therefore, if the size of the object and two of the 70, 100 and 160~$\mu$m fluxes are available, we recommend to measure the SFR surface density using the IR color adjusted version of Equation~\ref{eq:correlation_uncorr}:

\begin{gather}  
\Sigma_{\rm SFR}~({\rm M}_{\odot}~{\rm yr}^{-1}~{\rm kpc}^{-2}) = 3.79\times10^{-47} \nonumber \\
\times (\Sigma_{\rm[CII]}~[{\rm erg~s^{-1}~kpc^{-2}}]\times\Psi)^{1.13}. \nonumber
\end{gather}

\noindent where $\Psi$ is the color adjustment derived from Equation~\ref{eq:IRcorrection} and the values in Table~\ref{IRcolor_correction}.

\noindent Compared to pure starburst systems from the GOALS sample, this color-adjusted $\Sigma_{\rm [CII]}-\Sigma_{\rm SFR}$ correlation is valid over almost 5 orders of magnitude. We caution, however, that blind application of this calibration to systems that may host an AGN or where enough information to determine or bracket the color adjustment $\Psi$ is not available, risks a significant underestimate of the SFR.

For cases were no information on the size of the emitting object is available, we derive a SFR calibration (Equation~\ref{eq:correlation_lum}) based on the \cii\ luminosity, $L_{\rm [CII]}$. The dispersion in this correlation is similar to that of the $\Sigma_{\rm [CII]}-\Sigma_{\rm SFR}$ correlation. However, when compared to samples of galaxies with $L_{\rm TIR}>10^{11}~L_{\odot}$, our $L_{\rm [CII]}$--based calibration -- even after applying the IR color adjustment -- can underestimate the SFRs by more than a factor of $\sim3$. We suspect that the reason why the surface brightness calibration has better systematics is because it is more closely related to the local FUV field, most likely the main parameter controlling the photoelectric heating efficiency.

We highlight the following points:

\begin{enumerate}
\item We find a tight correlation between the surface brightness of \cii\ and 24~$\mu$m dust emission in normal galaxies.  Regions located in the central $\sim$1~kpc of galaxies that show AGN activity tend to show an excess of 24~$\mu$m emission compared to \cii. After excluding these points, the best linear fit yields a slope of $1.20\pm0.01$ \citep[which is close to the slope measured by][for the correlation between 24 $\mu$m and SFR]{rhc_calzetti07}. The scatter of the correlation is $\pm0.23$~dex.

\item For each individual region in our galaxy sample, we estimate the contribution from old stars to the 24~$\mu$m dust emission using the procedure described in Leroy et al. (2012). We refer to this emission as cirrus. We assume that the radiation field produced by these older populations is $U_{\rm cirrus}=0.5 \times U_{\rm min}$, where $ U_{\rm min}$ corresponds to the least interstellar radiation field heating the diffuse ISM in the \cite{rhc_draine07} model. We find that the median 24~$\mu$m cirrus contribution across the sample is 18\%. In order to obtain a more accurate measure of the SFR based on the 24~$\mu$m emission, we subtract from it our estimate of the cirrus contribution.

\item We estimate the SFR and $\Sigma_\text{SFR}$ values for each region using a combination of obscured (24~$\mu$m and TIR) and unobscured SFR tracers (H$\alpha$ and FUV). We then derive calibrations for the $\Sigma_\text{SFR}$ and SFR based on $L_{\rm[CII]}$ and $\Sigma_{\rm[CII]}$, respectively. The calibration coefficients can be found in Table~\ref{calibration_coeff}. We find that, for $\nu f_{\nu}(70)/\nu f_{\nu}(160)\gtrsim1.2$ or $\nu f_{\nu}(70)/\nu f_{\nu}(100)\gtrsim0.8$, the fit residuals systematically increase with increasing IR color (dust temperature), radiation field strength, fraction of the luminosity coming from regions with $U>100$ and decreasing $q_\text{PAH}$. At a slightly higher IR color threshold, $\nu f_{\nu}(70)/\nu f_{\nu}(100)\gtrsim0.95$, \cite{rhc_croxall12} find a drop in the \cii\ to FIR ratio for regions in NGC~4559 and NGC~1097. We parametrize the deviations we find for our warm regions as a function of IR color by a linear fit and derive a set of adjustments that reduces the residuals significantly. The list of IR color adjustments is given in Table~\ref{IRcolor_correction}.

\item For regions with oxygen abundances $12+{\rm log(O/H)}\lesssim8.1$, we find that our \cii-based SFR calibration is not reliable. Regions from the low metallicity galaxies Holmberg~II, IC~2574 and NGC~5408 show \cii-based SFRs that are a factor of $\sim4$ smaller than their SFR measured as a combination of 24~$\mu$m and FUV emission for the first two systems, and 24~$\mu$m for NGC~5408. These regions not only have the highest dust temperatures in the sample, but they also show significantly higher FUV to TIR ratios than their metal-rich counterparts. This suggests that a larger fraction of FUV photons escapes without interacting with the dust. If the \cii\ emission is mainly produced by grain photoelectric emission -- which requires FUV absorption by dust -- then the reduced trapping of FUV photons would explain the low \cii-based SFRs observed in low metallicity regions. In addition to this, we find that in NGC~5408 the \oi~$63\mu$m to \cii\ line ratio can be as high as $\sim1$, thus the cooling can be equally split between these two FIR transitions. For this particular galaxy, a SFR estimated from \cii\ alone will underpredict the total SFR by a factor of two. The line ratios and upper limits in Holmberg~II and IC~2574 do not rule out a similar scenario for these objects.

\item We find that an IR color adjusted $\Sigma_{\rm [CII]}$ can provide a good estimation of $\Sigma_{\rm SFR}$ using Equation~(\ref{eq:correlation_uncorr}), valid for starburst galaxies over almost 5 orders of magnitude in surface brightness. Without applying the IR color adjustment, KINGFISH regions and starbursts systems with $L_{\rm FIR}\leq10^{11}~L_{\odot}$ agree within a factor of $\sim2$ with Equation~\ref{eq:correlation_uncorr}. Starbursts with $L_{\rm FIR}\geq10^{11}~L_{\odot}$ tend to follow the same trend, but with $\Sigma_\text{SFR}$ values that are, on average, a factor of $\sim3$ higher for a given $\Sigma_{\rm[CII]}$. 

\item In the luminosity regime, the SFR calibration defined by Equation~\ref{eq:correlation_lum} works well for samples of normal, star-forming galaxies \citep{rhc_boselli02,rhc_delooze11}, but underestimates the SFR derived from the TIR value by a factor greater than $\sim3$ for more than half of the GOALS galaxies and non-AGN LIRGs (Sargsyan et al. 2012; Weiner et al. in prep.). The IR color adjustment helps to reduce the discrepancy -- especially for the GOALS sample -- but even after applying the adjustment there are LIRGs that show deviations as high as a factor $\sim$10. This demonstrates the limitations of using the \cii\ luminosity as a SFR measure in warm or compact IR luminous and ultra luminous galaxies,  for which low [CII] to TIR ratios -- or the so called ``\cii-deficit'' -- have been extensively reported in the literature. One additional factor behind these deviations is the different tracers we are using to measure the SFR of LIRGs (TIR) and the KINGFISH regions (24~$\mu$m combined with H${\alpha}$ or FUV). Interestingly, we find that if we measure the SFR of LIRGs using monochromatic IR-based SFR tracers (e.g., 24~$\mu$m), the agreement with the SFR inferred from \cii\ through our calibration is considerably better.

\item We use the {\it Starburst99} code to connect the FUV luminosity of modeled stellar populations to the \cii\ emission via the heating efficiency, $\epsilon_{h}$. We find that the \cii\ emission from most of the galaxies can be attributed to regions that have been forming stars continuously for more than 20~Myr in combination with a heating efficiency in the range $\epsilon_{h}\sim1-3\%$. It appears likely that the variation in the latter drives much of the scatter in the \cii--SFR correlation.
\end{enumerate}

We thank the referee for constructive and valuable comments on the paper. R.H.C wishes to acknowledge support from a Fulbright-CONICYT grant. A.D.B. wishes to acknowledge partial support from a CAREER grant NSF-AST0955836, from NSF-AST1139998, from NASA-JPL 1373858 and from a Research Corporation for Science Advancement Cottrell Scholar award. We thank
B. Weiner for providing \cii\ and FIR fluxes for his sample of LIRGs in advance of publication. We also thank T. Diaz-Santos for providing the areas used to measure the surface brightness of the galaxies in the GOALS sample in advance of publication. PACS has been developed by a consortium of institutes led by MPE (Germany) and including UVIE (Austria); KU Leuven, CSL, IMEC (Belgium); CEA, LAM (France); MPIA (Germany); INAF-IFSI/OAA/OAP/OAT, LENS, SISSA (Italy); IAC (Spain). This development has been supported by the funding agencies BMVIT (Austria), ESA-PRODEX (Belgium), CEA/CNES (France), DLR (Germany), ASI/INAF (Italy), and CICYT/MCYT (Spain). HIPE is a joint development by the Herschel Science Ground Segment Consortium, consisting of ESA, the NASA Herschel Science Center, and the HIFI, PACS, and SPIRE consortia. This work is based (in part) on observations made with Herschel, a European Space Agency Cornerstone Mission with significant participation by NASA. The National Radio Astronomy Observatory is a facility of the National Science Foundation operated under cooperative agreement by Associated Universities, Inc. This research has made use of the NASA/IPAC Extragalactic Database (NED) which is operated by the Jet Propulsion Laboratory, California Institute of Technology, under contract with the National Aeronautics and Space Administration.

We wish to dedicate this study to the memory of our colleague Dr. Charles ``Chad" Engelbracht for his numerous contributions to space-based far-infrared astronomy that made this work possible, as well as that of many other researchers employing {\it Spitzer} and {\it Herschel} observations.

\bibliography{references.bib}

\newpage


\begin{appendix}

\section{KINGFISH galaxies not included in the analysis}

Among the 54 galaxies in the KINGFISH sample with PACS spectroscopic data available, we decided not to include eight of them: NGC~1266, NGC~1316, NGC~1097, NGC~1377, NGC~1404, NGC~4594, NGC~4631 and NGC~4559. We excluded the two elliptical galaxies NGC~1266 and NGC~1316 (also know as Fornax~A) because they both have AGNs \citep{rhc_ekers83, rhc_moustakas10, rhc_nyland13} that may contaminate the infrared emission used to measure SFRs. Unfortunately, masking the AGN emission at the center of these galaxies is not an option because it removes a significant fraction of the \cii\ map. NGC~1377 is a peculiar system that shows a  24~$\mu$m -- \cii\ ratio nearly two orders of magnitude higher than the rest of the points at a given \cii\ surface brightness. The strong infrared excess in this system is produced either by a nascent starburst \citep{rhc_roussel06} or a buried AGN \citep{rhc_imanishi09}. Given that the central source that dominates the emission of NGC~1377 is debated, we remove this system from the analysis.  We also do not include NGC~1404 and NGC~4594 because the quality of the spectroscopic data is poor. We remove NGC~4631 from the sample because of its high inclination of 86$^{\circ}$ \citep{rhc_mmateos09}. Finally, we do not include NGC~1097 and NGC~4559 because these two galaxies were observed in the Science Demonstration Phase using a different observing mode (i.e. Chop-Nod and Wavelength Switching) than the one used to observe the rest of the sample (Unchopped Grating Scan). Note, however, that these two systems agree well with the correlations presented in this work.

\section{Supplementary data}

{\it SINGS \& KINGFISH IR:} For all galaxies in our sample we have images that cover the entire IR continuum from 3.6~$\mu$m to 500~$\mu$m. Near and mid-infrared (8 and 24~$\mu$m) images were drawn from the {\it Spitzer} Infrared Nearby Galaxy Survey \citep[SINGS]{rhc_kennicutt03}. Far-infrared maps observed with {\it Herschel} PACS (70, 100  and 160~$\mu$m) and SPIRE (250, 350 and 500~$\mu$m) instruments are drawn from the photometric KINGFISH sample \citep{rhc_dale12}. We also add \oi~63~$\mu$m data reduced and calibrated in the same way as the \cii\ data. \\

{\it H$\alpha$:} We have H$\alpha$ images for 27 galaxies. The assembly of these images and the methods to correct for Galactic extinction, mask foreground stars and remove the [NII] contribution are described in detail in \cite{rhc_leroy12}. The source of these images are (in order of importance): SINGS \citep{rhc_kennicutt03}, Local Volume Legacy survey \citep[LVL,][]{rhc_dale09}, GOLDMine \citep{rhc_gavazzi03}, Palomar Las Campanas Atlas, \cite{rhc_boselli02b}, \cite{rhc_knapen04} and \cite{rhc_hoopes01}. \\

{\it FUV:} We have GALEX FUV images for 33 of the galaxies in our sample. The assembly of these images and the additional processing that includes background subtraction and masking of foreground stars via their UV color, by-eye inspection, and the color-based masks of \cite{rhc_mmateos09b} is described in \cite{rhc_leroy12}. The source of these images are (in order of importance): Nearby Galaxy Survey \citep[NGS,][]{rhc_gdepaz07}, Medium Imaging Survey (MIS) and All-sky Imaging Survey \citep[AIS,][]{rhc_martin05}. \\

{\it Draine \& Li Dust Model Maps:} We use maps of dust properties, like the ones presented in \cite{rhc_aniano12}, based on the Draine \& Li dust model \citep{rhc_draine07} (DL07). The DL07 model treats the dust as a combination of carbonaceous and amorphous silicate grains, with the smallest carbonaceous grains having the physical properties of PAH particles. The PAH mass fraction, $q_\text{PAH}$, is defined as the percentage of the total grain mass contributed by PAHs containing fewer than 10$^{3}$~C atoms \citep[see][eq. 4]{rhc_draine07b}. The grain size distribution and normalization is chosen to match the abundance and average extinction in the Milky Way. The DL07 model considers that dust is exposed to a range of radiation fields: a ``diffuse ISM" component -- that contains most of the dust -- which is heated by a single ($\delta$ function) radiation field, $U = U_\text{min}$; and a ``PDR component" that is heated by a power-law distribution of intensities $U$ over a wide range, $U_{\rm min} < U < U_\text{max}$, where ($U_\text{max} \gg U_\text{min}$). For the ``PDR component", the DL07 model estimate the fraction of the dust luminosity radiated from regions where $U > 100$ \citep[see][eq. 18]{rhc_draine07b}. Finally, the DL07 model also yields a dust mass ($M_\text{dust}$) and a dust mass surface density ($\Sigma_{\rm dust}$) for each line of sight. In this paper we use the version of the dust maps generated by fitting the IR SED composed of the MIPS~24~$\mu$m and PACS~70, 100 and 160~$\mu$m fluxes. We use this version of the dust maps, as opposed to the version that include SPIRE fluxes, to have dust maps with similar spatial resolution to the \cii~158~$\mu$m maps. 

\section{Cirrus Emission}

Following the procedure described in \cite{rhc_leroy12}, the 24~$\mu$m cirrus intensity ($I_{24}^{\rm cirrus}$) can be computed as the product of the dust surface density ($\Sigma_{\rm dust}$), and the emission per unit dust mass of dust heated by non star-forming sources ($\epsilon_{24}^{\rm cirrus}$):

\begin{equation}
I_{24}^{\rm cirrus} = \epsilon_{24}^{\rm cirrus}(U_{\rm cirrus},q_{\rm PAH}) \times \Sigma_{\rm dust}.
\end{equation}

\noindent In the \cite{rhc_draine07} model, $\epsilon_{24}$ depends linearly on the incident radiation field, $U$ ($U=1$ is the local interstellar radiation field in the Milky Way, normalized to the MMP field \citep{rhc_mathis83}) and weakly on the PAH abundance index, $q_{\rm PAH}$ (Fig. 4, Leroy et al. 2012). For instance, for a radiation field $U=1$, the emission per unit dust mass $\epsilon_{24}$ increases only by a factor of $\sim1.75$ in the $q_{\rm PAH}=0.47-4.58\%$ range \citep[Table 4]{rhc_draine07}. Therefore, the challenge here is to find the incident radiation field produced by non star-forming sources, i.e., $U_\text{cirrus}$. \cite{rhc_leroy12} test different values of $U_\text{cirrus}$ with the goal of producing 24~$\mu$m cirrus emission such that its subtraction removes the 24~$\mu$m faint emission, but not oversubtract emission associated with star formation. For a sample of 30 disk galaxies (20 of them part of the KINGFISH sample), they find that  this optimal cirrus radiation field is $U_\text{cirrus} \approx 0.5~U_\text{min}$, where $U_\text{min}$ corresponds - in the \cite{rhc_draine07} model - to the least interstellar radiation field heating the diffuse ISM. There is a factor of $\sim2$ scatter associated with this value \citep[Appendix A]{rhc_leroy12}.

We explore the effects of the cirrus subtraction estimating $U_\text{cirrus}$ as a scaled version of $U_\text{min}$, i.e., $U_\text{cirrus} = \xi \times U_\text{min}$ for $\xi =$ 0.5, 0.75 and 1. We also explore adopting a constant $U_\text{cirrus}$ across the galaxy. Given that we measure a median value of $U_\text{min}$ in our sample of 1.15, we adopt a constant $U_\text{cirrus} = \xi \times 1.15 = 0.6, 0.8$ and 1.1. Choosing a high cirrus radiation field, such as $U_\text{cirrus} = U_\text{min}$ or $U_\text{cirrus} = 1.1$, will likely overestimate the cirrus emission. As \cite{rhc_leroy12} point out, the OB associations present within 1~kpc of the Sun would substantially contribute to the local interstellar radiation field; so non star-forming sources - like old stars - cannot by themselves account on their own to the local measured value of $U=1$. 

Table~\ref{table1} summarizes the effect of the cirrus correction on the $\Sigma_{24\mu\text{m}}$--$\Sigma_{\rm [CII]}$ correlation and the fraction of the 24~$\mu$m emission associated with cirrus ($f_\text{cir}$). When cirrus is estimated using $U_\text{cirrus} = 0.5~U_\text{min}$, the shape and scatter of the corrected correlation remains essentially the same, but the normalization scales down about 18\% (as expected since $f_\text{cir}\sim18\%$ for this case). For the cirrus removal based on $U_\text{cirrus} = 0.6$, the cirrus subtraction is higher at lower luminosities of $24~\mu$m, thus we observe an increase in the slope from 1.20 to 1.30; the scatter of the correlation remains the same. For these two sets of assumptions on $U_\text{cirrus}$, the median $f_\text{cir}$ is similar to the $f_\text{cir}\sim$19\% found by \cite{rhc_leroy12} (the overlap between our samples is 23 galaxies) and larger than the $f_\text{cir}\sim$7\% found by \cite{rhc_law11} in their analysis of integrated SEDs for LVL and SINGS galaxies. For the $U_\text{cirrus}=1.1$ case, the changes in the correlation after the cirrus removal are more dramatic: the slope changes from 1.20 to 1.60; as a result, the subtracted 24~$\mu$m emission in the low \cii\ luminosity regions ($\Sigma_{\rm[CII]}<10^{39}$~[erg~s$^{-1}$~kpc$^{-2}$]) can be a factor of $\sim2$ smaller compared to the uncorrected correlation. 

\section{Dust Attenuation}

Another physical variable of interest in the analysis of the local variations is the dust attenuation. For 33 galaxies for which we have FUV images available, we measure the fraction of FUV and optical emission absorbed and reprocessed by dust versus the escaping FUV emission as the ratio of the 24~$\mu$m to the FUV intensity, $I_{24\mu\rm{m}}/I_{\rm{FUV}}$. Figure~\ref{residual_extinction} shows the residuals of the correlation between the \cii-based $\Sigma_\text{SFR}$ and $\Sigma_\text{SFR}({\rm FUV}+24~\mu{\rm m})$. The top panel shows the fit residuals for all the 12\arcsec regions from the 33 galaxies for which we have FUV maps available. Around the $I_{24\mu\rm{m}}/I_{\rm{FUV}}$ ratio of $\sim10$, we observe a systematic increase of the fit residuals as a function of decreasing extinction. The bottom panel of Figure~\ref{residual_extinction} shows that the regions that deviate are mainly coming from two galaxies: NGC~6946 and NGC~7793. Visual inspection of the \cii\ and dust attenuation maps reveals that the regions with low $I_{24\mu\rm{m}}/I_{\rm{FUV}}$ ratios are located preferentially in regions of low surface brightness in the ``extranuclear'' pointings. The rest of the regions that show high fit residuals at low $I_{24\mu\rm{m}}$ to $I_{\rm{FUV}}$ ratios are from the low metallicity systems (12+log$_{10}$(O/H)$<8.1$) Holmberg~II and IC~2574. 

\begin{figure}
\epsscale{0.6}
\plotone{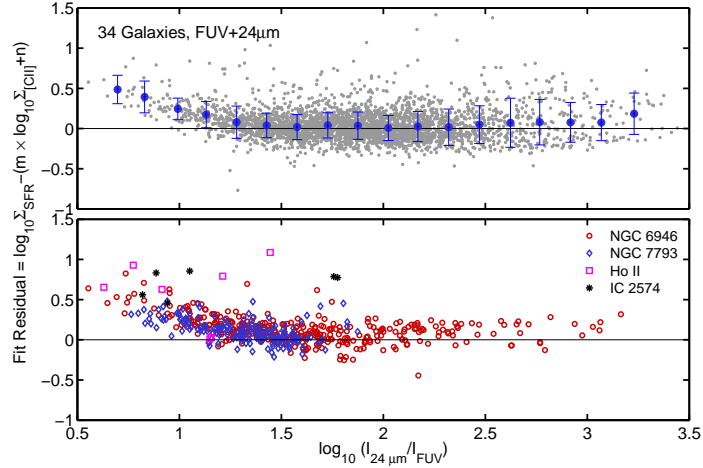}
\caption{Fit residual of the $\Sigma_{\rm[CII]} - \Sigma_\text{SFR}$ correlation as a function of the  fraction of FUV absorbed and reprocessed by dust versus the escaping FUV emission ($I_{24\mu\rm{m}}/I_{\rm{FUV}}$). Each point represent a 12\arcsec size region. Blue dots and the corresponding vertical bars correspond to the median and the 1~$\sigma$ standard deviation of the binned distribution of points. {\it (Top panel)} For 33 galaxies for which we have FUV maps available, we measure $\Sigma_\text{SFR}$ as a linear of combination of 24~$\mu$m and FUV emission \citep{rhc_leroy08} and the dust attenuation as the ratio of the 24~$\mu$m intensity, $I_{24~\mu\rm{m}}$ [MJy~sr$^{-1}$], and the FUV intensity, $I_{\rm{FUV}}$ [MJy~sr$^{-1}$]. {\it (Bottom panel)} Four individual galaxy cases: two galaxies with global average low metallicity (Holmberg~II and IC~2574) and two galaxies that are the main source of the low extinction regions that deviate from the fit (NGC~6946 and NGC~7793). \label{residual_extinction}}
\end{figure}

\section{Additional samples of galaxies included in our analysis}

{\it \cite{rhc_diaz-santos13}}: From the 241 systems comprising the GOALS sample, we only select a subset of 116 systems that are classified as pure starburst following the criteria described in \cite{rhc_diaz-santos13}, i.e., that have $6.2~\mu$m PAH equivalent widths that are greater than 0.5. The sample covers a distance range of $\sim16-350$~Mpc. We compiled PACS \cii\ flux densities, \cii\ to FIR ratios and PACS continuum flux densities at 63 and 158~$\mu$m under the \oi\ and \cii\ lines, respectively, from \cite{rhc_diaz-santos13}. The areas of the infrared emitting nuclear regions of the GOALS sample were provided by T. Diaz-Santos (priv. comm.).

The FIR luminosities in \cite{rhc_diaz-santos13} were measured based on the 60 and 100~$\mu$m IRAS bands using \cite{rhc_helou88} calibration. In order to use the SFR calibration by \cite{rhc_murphy11} (or \cite{rhc_kennicutt98}, which yields a SFR value $\sim 1.3$ times higher), we need to convert these FIR($60-100~\mu$m) luminosities into TIR($8-1000~\mu$m) luminosities. The ideal would be to measure the TIR luminosity of the GOALS galaxies by integrating their SED from 8 to $1000~\mu$m. This can be done for 64 U/LIRGs from the GOALS sample, for which \cite{rhc_vu12} compiled aperture photometry from radio through X-ray wavelengths. Based on these SEDs, \cite{rhc_vu12} use different dust models and a modified blackbody fit to measure the TIR luminosity from 8 to $1000~\mu$m. They find that these integrated TIR($8-1000~\mu$m) luminosities are about 0.02~dex lower than the FIR($12-100~\mu$m) luminosities measured using the calibration by \cite{rhc_sanders96}. Compared to the FIR($60-100~\mu$m) luminosities based on the \cite{rhc_helou88} calibration, the TIR($8-1000~\mu$m) luminosities are higher by a factor of 1.74. We adopt this factor to convert the FIR luminosities listed in \cite{rhc_diaz-santos13} into TIR($8-1000~\mu$m) luminosities, and then we measure the SFR based on TIR($8-1000~\mu$m) using the calibration by \cite{rhc_murphy11}. \\

{\it \cite{rhc_boselli02}}: This work encompasses 23 galaxies detected in \cii\ line emission by ISO. The sample include 18 spiral systems from the Virgo cluster ($D=17$~Mpc). The \cii, H$\alpha$ and FIR luminosities were taken from the paper. Additional IRAS~60 and 100~$\mu$m fluxes were added from \cite{rhc_leech99} and \cite{rhc_smith97}. We measure the SFR from the H$\alpha$ luminosities using the \cite{rhc_calzetti07} calibration. In order to derive individual IR color adjustments, we interpolate the value of the 70~$\mu$m flux based in the 60 and 100~$\mu$m fluxes, and then we measure the 70/100~$\mu$m ratio for each galaxy.\\

{\it \cite{rhc_delooze11}:} This work encompasses 17 star-forming and starburst galaxies observed by ISO and located within 60~Mpc, with the exception of one system that is 139~Mpc away. The \cii, 24~$\mu$m, FUV and FIR luminosities were taken from the paper. We measure the SFR as a combination of 24~$\mu$m and FUV emission using the \cite{rhc_leroy08} calibration. For the IR color adjustment, we compile 60 and 100~$\mu$m fluxes from \cite{rhc_brauher08}. Based on interpolated values of the 70~$\mu$m flux, we measure the 70/100~$\mu$m ratio for each galaxy in the sample.\\

{\it \cite{rhc_sargsyan12}:} This work includes 23 starbursting LIRGs observed by {\it Herschel} and located in the distance range from $\sim$66 to 505~Mpc. We compiled \cii\ and FIR($12-100~\mu$m) luminosities from their paper. As we mentioned in \S4.2, the FIR($12-100~\mu$m) luminosities measured using \cite{rhc_sanders96} calibration are similar to the integrated TIR($8-1000~\mu$m) luminosities in U/LIRGs \citep{rhc_vu12}. Therefore, we use the FIR($12-100~\mu$m) luminosities to measure the SFR based on \cite{rhc_murphy11} calibration. \\

{\it Weiner et al. in prep.:} This work includes 16 disky non-mergers, non-AGN, LIRGs located at  $z\sim0.1$. The \cii\ 158~$\mu$m, 60 and 100~$\mu$m IRAS fluxes were provided by B. Weiner (priv. comm.). We measure the FIR luminosities using the 60 and 100~$\mu$m fluxes. We then measure the TIR luminosities, SFRs and 70/100~$\mu$m IR colors using the same procedure applied to the \cite{rhc_diaz-santos13} sample.\\

\newpage

\section{$\Sigma_{\rm [CII]} - \Sigma_{24~\mu\rm{m}}$ Correlation for the KINGFISH Galaxies}

\begin{figure*}[h]
\epsscale{1}
\plotone{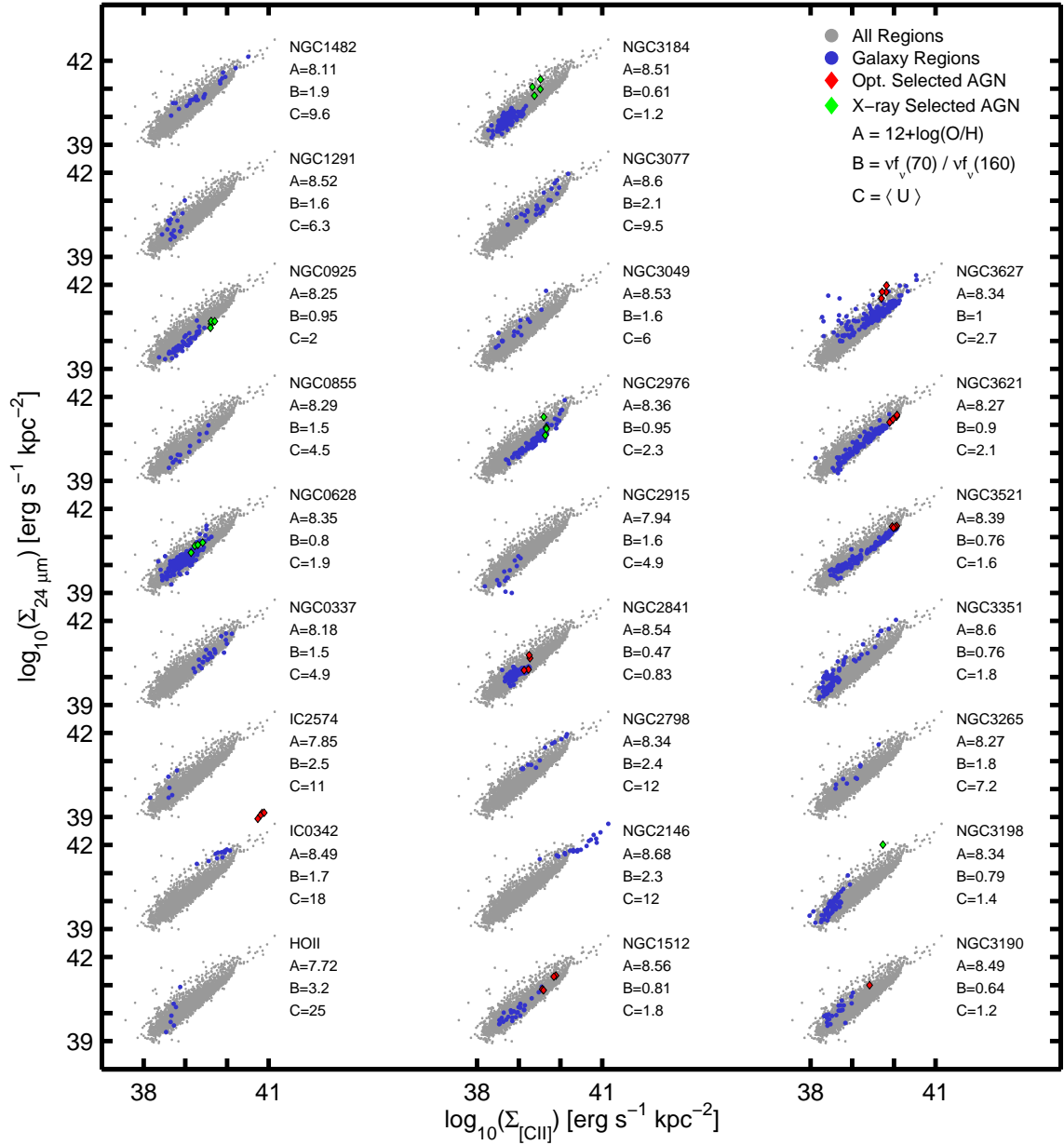}
\caption{[CII] luminosity surface density versus 24~$\mu$m surface density for each of the KINGFISH galaxies in our sample. The grey dots correspond to all the regions from the 46 KINGFISH galaxies in our sample. The blue dots correspond to the regions of the galaxy whose name is on the top right of the correlation. We also list the metallicity (A), mean IR color (B) and mean starlight intensity of the galaxy. If the galaxy host an AGN, the regions from the central $\sim0.5$~kpc radius are show as diamonds: red for optically selected AGN \citep{rhc_moustakas10} and green for X-ray selected AGN \citep{rhc_tajer05, rhc_grier11}.\label{cii_panel}}
\end{figure*}

\begin{figure*}[h]
\epsscale{1}
\plotone{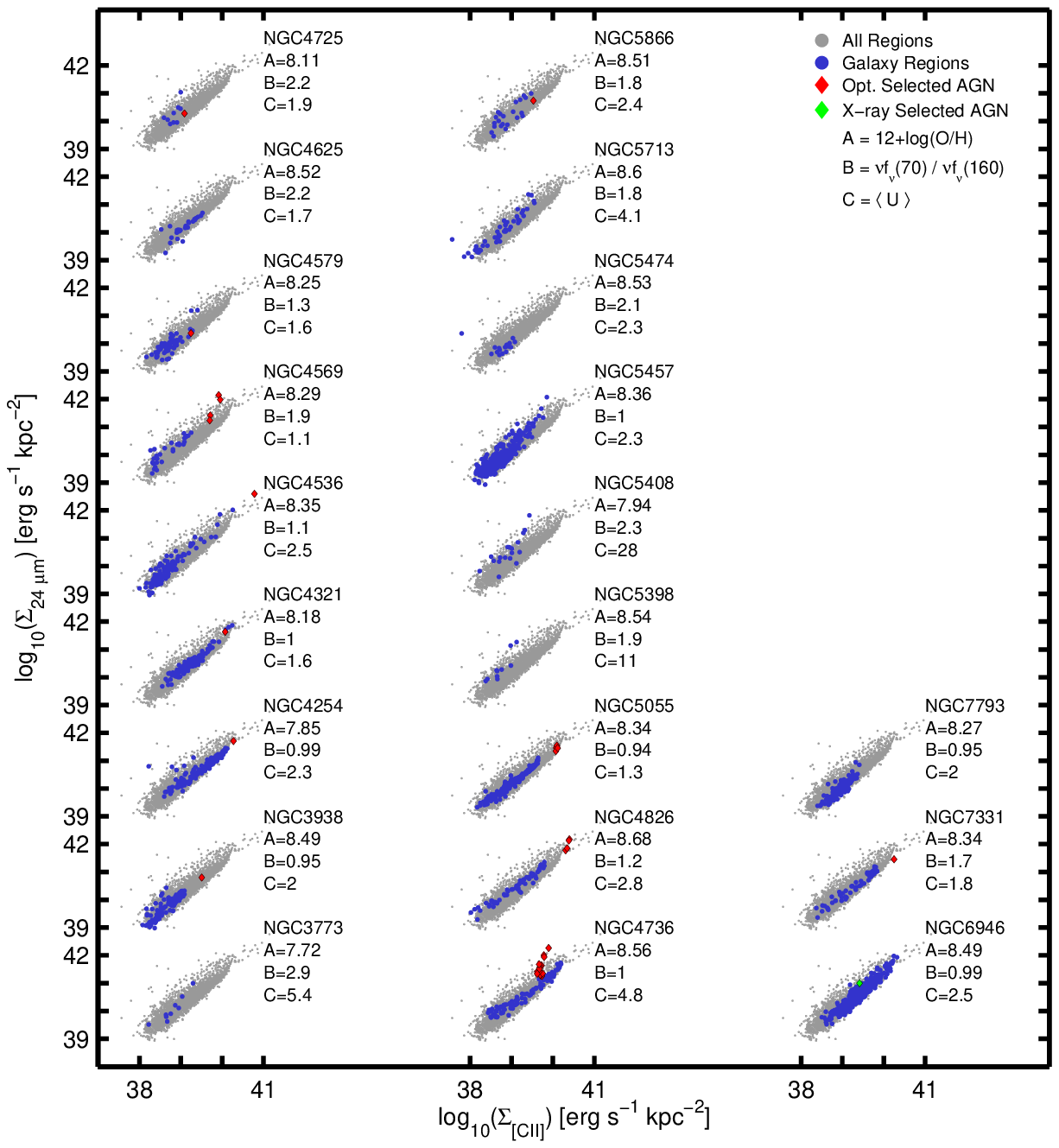}
\caption{As Figure~\ref{cii_panel}. \label{cii_panel2}}
\end{figure*}

\end{appendix}

\end{document}